\documentclass[pra,twocolumn,superscriptaddress,pdfoutput=1]{revtex4-1}
\usepackage{graphicx}
\usepackage{epsfig}
\usepackage{amssymb,amsmath,amsfonts}
\usepackage{wasysym}
\usepackage{bm}
\usepackage{latexsym}
\usepackage{eucal}
\usepackage{float}
\usepackage{verbatim}
\usepackage[normalem]{ulem}
\usepackage{epstopdf}
\usepackage{color}
\usepackage[dvipsnames]{xcolor}
\usepackage{braket}
\usepackage{mathrsfs}
\usepackage[colorlinks,citecolor=blue]{hyperref}

\begin{document}

\title{$\mathbb{Z}_{2}$ topological order and first-order quantum phase transitions \\ in systems with combinatorial gauge symmetry}

\author{Kai-Hsin Wu}
\email{khwu@bu.edu}
\affiliation{Department of Physics, Boston University, 590 Commonwealth Avenue, Boston, Massachusetts 02215, USA}

\author{Zhi-Cheng Yang}
\email{zcyang@umd.edu}
\affiliation{Joint Quantum Institute, University of Maryland, College Park, MD 20742, USA}
\affiliation{Joint Center for Quantum Information and Computer Science, University of Maryland, College Park, Maryland 20742, USA}

\author{Dmitry Green}
\email{dmitry.green@aya.yale.edu}
\affiliation{AppliedTQC.com, ResearchPULSE LLC, New York, NY 10065, USA}

\author{Anders W. Sandvik}
\email{sandvik@bu.edu}
\affiliation{Department of Physics, Boston University, 590 Commonwealth Avenue, Boston, Massachusetts 02215, USA}
\affiliation{Beijing National Laboratory for Condensed Matter Physics and Institute of Physics, Chinese Academy of Sciences, Beijing 100190, China}

\author{Claudio Chamon}
\email{chamon@bu.edu}
\affiliation{Department of Physics, Boston University, 590 Commonwealth Avenue, Boston, Massachusetts 02215, USA}
	
\date{\today}

\begin{abstract}
We study a generalization of the two-dimensional transverse-field
Ising model, combining both ferromagnetic and antiferromagnetic
two-body interactions, that hosts exact global and local
$\mathbb{Z}_2$ gauge symmetries. Using exact diagonalization and
stochastic series expansion quantum Monte Carlo methods, we confirm
the existence of the topological phase in line with previous
theoretical predictions. Our simulation results
show that the transition between the confined topological phase and
the deconfined paramagnetic phase is of first-order, in contrast to
the conventional $\mathbb{Z}_2$ lattice gauge model in which the
transition maps onto that of the standard Ising model and is continuous. We
further generalize the model by replacing the transverse field on the
gauge spins with a ferromagnetic $XX$ interaction while keeping the local
gauge symmetry intact. We find that the $\mathbb{Z}_2$ topological
phase remains stable, while the paramagnetic phase is replaced by a
ferromagnetic phase. The topological--ferromagnetic quantum phase
transition is also of first-order. For both models, we discuss the
low-energy spinon and vison excitations of the topological phase and
their avoided level crossings associated with the first-order quantum
phase transitions.
\end{abstract}

\maketitle

\section{Introduction}
Topological quantum states of matter are of central focus in modern
condensed matter physics. One of the main features of
strongly-interacting systems with gapped topological order is that
they can present degenerate ground states. This degeneracy cannot be
lifted by the action of local perturbations, and hence this property
makes such systems perfect candidates for building stable
(topological) qubits. Several theoretical models have been proposed to
realize gapped topologically ordered states. For instance, the toric
code~\cite{KITAEV20032} and dimer models on non-bipartite lattices
\cite{MoessnerRVB2001,KagomeRVB2002} host quantum spin liquid (QSL)
phases that possess $\mathbb{Z}_2$ topological order. In both these
examples, the Hamiltonians contain multi-body interactions, making it
a challenge to encounter materials realizing these phases or to
realize them in artificial structures.
	
Attempts have been made to construct models with simpler interactions
that can host gapped QSLs. A rare successful example is the cluster
charging model of bosons on the kagome
lattice~\cite{BalentsXXZ2002,IsakovXXZ2006,YBXXZ2007}, which has been
shown theoretically and numerically to host a $\mathbb{Z}_2$ quantum
spin liquid, in a system with only two-body interactions. However,
these two-body interactions are of the $XXZ$ type, which are not easily
implementable in, say, programmable quantum devices. Moreover, the
$\mathbb{Z}_2$ gauge symmetry in this model is only emerging, i.e., it
exists in the effective model derived in perturbation theory, but it
is not an exact symmetry of the original Hamiltonian.
	
Recently, a construction for which the $\mathbb{Z}_2$ gauge symmetry
is exact was proposed on a variant of the transverse field Ising model
(TFIM), utilizing only simple two-body ferromagnetic and
antiferromagnetic $ZZ$ interactions~\cite{chamon2019emulate}.
Monomial (matrix) transformations that correspond to combinations of
spin flips and permutations play a central role in the construction,
thus dubbed combinatorial gauge symmetry.  Because the construction
utilizes only $ZZ$ interactions (of both signs) and a transverse
field, the model can be easily implemented, for example, with current
Noisy Intermediate-Scale Quantum (NISQ) technology using flux-based
superconducting qubits, or other types of quantum computer
architectures that provide similar interactions on qubits. The model
has already been successfully implemented on a D-wave quantum device
in a recent experiment~\cite{zhou2020building}.
	
In this paper, we present a quantitative and detailed study of the
combinatorial $\mathbb{Z}_2$ gauge model originally proposed in
Ref.~\onlinecite{chamon2019emulate}.
Two different types of quantum fluctuations are introduced while
preserving the gauge symmetry: a transverse field acting on the gauge
spins and a $XX$ ferromagnetic interaction between the gauge spins,
respectively. In both cases, we observe the existence of a
$\mathbb{Z}_2$ topological state separated by a first-order transition
from another phase---a paramagnet in the first model and a
ferromagnet in the second

The structure of the paper is as follows. First, in Sec.~\ref{sec:combinatorial}
we give a brief introduction to the model with combinatorial $\mathbb{Z}_2$
gauge symmetry. The model realizes the $\mathbb{Z}_2$ gauge symmetry through
monomial transformations and effectively realizes the 4-body interaction term as
the star term in the classical version of the toric code. We further introduce
two types of quantum fluctuations by applying either a transverse field on the
gauge spins (model-X) or a $XX$ ferromagnetic interaction between the gauge
spins (model-XX), both of which respect the gauge symmetry. In
Secs.~\ref{sec:tfimgauge} and~\ref{sec:numerics_XX}, we provide
numerical results on both models obtained from quantum Monte-Carlo
(QMC) simulations with the Stochastic Series Expansion (SSE) method
as well as exact diagonalization (ED). In both cases, we find that the system
exhibits a $\mathbb{Z}_2$ topologically ordered phase separated by a first-order
transition from either a paramagnetic phase (model-X) or a ferromagnetic
phase (model-XX). We summarize our main results and discuss the remaining open
questions and future prospects in Sec.~\ref{sec:conclusion}.

\begin{figure}[t]
\includegraphics[width=75mm]{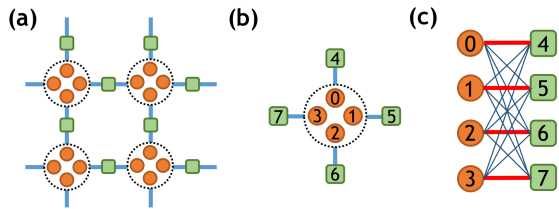}
\caption{(a) The arrangement of two species of Ising spins on a square
  lattice. Gauge spins (green squares) reside on the links and four
  matter spins (orange dots) occupy each site of the square
  lattice. (b) The vertex unit and the interactions between spins.
  The interactions between gauge-matter spins within a single unit are
  defined by the Hadamard matrix $\mathbf{W}$ in Eq.~(\ref{eq.W}). For clarity,
  the couplings are depicted on a deformed cell in panel (c), where the
  ferromagnetic and antiferromagnetic couplings are shown as thin blue lines
  and bold red lines, respectively.}
\label{fig.geo}
\end{figure}

\section{Spin models with combinatorial gauge symmetry}
\label{sec:combinatorial}

We consider two models in our studies starting from a baseline model
that contains $ZZ$ interactions between the spins with both ferromagnetic
and antiferromagnetic couplings arranged in a pattern such that the
combinatorial gauge symmetry is realized.

On top of the baseline model, additional kinetic terms are introduced
to give the system quantum dynamics. Here we consider two different
types of kinetic terms. The first one is a transverse field on the gauge
spins in what we call model-X; the second one is a ferromagnetic XX-type
($\sigma^x\sigma^x$) coupling between the gauge spins, which defines model-XX. In this
section we start by describing the details of the baseline model and how the
combinatorial gauge symmetry is realized. We then discuss the models with
the two different kinetic terms.

\subsection{Baseline model with combinatorial gauge symmetry}

In the baseline model $S=1/2$ spins reside on both the sites and the links of
a square lattice, as shown in Fig.~\ref{fig.geo}(a). For each star (or vertex)
$s$ of the lattice, we place four ``matter'' spins $\mu$ (the orange dots inside
the circles representing the lattice sites) and four ``gauge'' spins $\sigma$
(green square) on the links of the star. One such star with its total of eight
spins is depicted in Fig.~\ref{fig.geo}(b) along with a labeling scheme.
The Hamiltonian of the system is written as
\begin{align}
  H = \sum_s H_s + H^\sigma_{\rm kin}
  \;,
  \label{eq:Hgeneral}
\end{align} 
where $H^\sigma_{\rm kin}$ is the kinetic term involving only the
$\sigma^x$ component of the gauge spins (on all links), and at each
star $s$ we define a local Hamiltonian on its eight spins
\begin{align}
  H_s
  =
  J\sum_{a\in s} \left(\sum_{j\in s} W_{aj} \sigma^z_j \right) \mu^z_a
  -
  \Gamma_m \sum_{a\in s} \mu^x_a
  \;.
  \label{eq.H0}
\end{align}
Here the sums with $j$ and $a$ indices include the four
matter spins $\sigma^z_j$ and the four gauge spins $\mu^z_a$,
respectively. Notice that $H_s$ contains a transverse field
only on the matter spins. The $ZZ$ couplings between the gauge and
matter spins have magnitude $J$ and signs controlled by the Hadamard
matrix $W$,
\begin{align}
\mathbf{W} = 
\begin{pmatrix}
	+1 & -1 & -1 & -1 \\
	-1 & +1 & -1 & -1 \\
	-1 & -1 & +1 & -1 \\
	-1 & -1 & -1 & +1
\end{pmatrix}. \label{eq.W}
\end{align}
Fig.~\ref{fig.geo}(c) depicts these signs; the interaction between a
gauge spin and its nearest matter spin is antiferromagnetic (bold red
line) while its interaction with the other 3 matter spins are
ferromagnetic (thin blue line).

The Hamiltonian in Eq.~\eqref{eq:Hgeneral} possesses combinatorial
$\mathbb{Z}_2$ gauge symmetry if the $H_s$ terms are of the above
form, and if only the $\sigma^x$ component of the gauge spins enters
in $H^\sigma_{\rm kin}$. The transformations
\begin{subequations}
\begin{align}
\sigma^{z}_i &\rightarrow \sum_j \mathbf{R}_{ij}\,\sigma_j^{z}  \\
\mu^{z}_b &\rightarrow \sum_a \mu_a^{z}\, \mathbf{L}^{-1}_{ab}
\end{align}
\label{eq:RL-trans}
\end{subequations}
leave the spin commutation relations invariant if $\mathbf{L}$ and $\mathbf{R}$ are
monomial matrices, i.e., generalized permutation matrices with a $\pm 1$ entry in each
line or column in the case of the group $\mathbb{Z}_2$. The transformations correspond
to combinations of rotations by 0 (+1 entry) or $\pi$ (-1 entry) around the $x$-axis,
followed by a permutation of the indices. 

A local gauge symmetry is generated by flipping gauge spins on closed loops around the elementary plaquettes, together with accompanying transformations on matter spins. 
Flipping the gauge spins around the loop corresponds to choosing $\mathbf{R}$ matrices for each $s$ traversed, with an even number of $-1$ entries associated with the links visited. 
For each such $\mathbf{R}$, there is a corresponding monomial matrix $\mathbf{L}=\mathbf{W}\,\mathbf{R}\,\mathbf{W}^{-1}$~\cite{chamon2019emulate}. 
These pairs of monomial $\mathbf{R}$ and $\mathbf{L}$ matrices are such that $\mathbf{W}=\mathbf{L}^{-1}\,\mathbf{W}\,\mathbf{R}$, and thus the transformation in Eq.~\eqref{eq:RL-trans} leaves the $ZZ$ part of the Hamiltonian invariant. 
Moreover, since in $H_s$ the transverse field on the $\mu^x$ is the same on all $a\in s$, the permutation action of the monomial $\mathbf{L}$ also leaves these terms unchanged. 
Hence, $H_s$ is invariant under the monomial transformation with $\mathbf{L}$ and $\mathbf{R}$. Finally, since $\mathbf{R}$ is diagonal and only $\sigma^x$ enters in $H^\sigma_{\rm kin}$, this kinetic term is also invariant. For each such loop, we have a $\mathbb{Z}_2$ symmetry of the Hamiltonian in Eq.~\eqref{eq:Hgeneral}.

\begin{figure}[t]
\includegraphics[width=75mm]{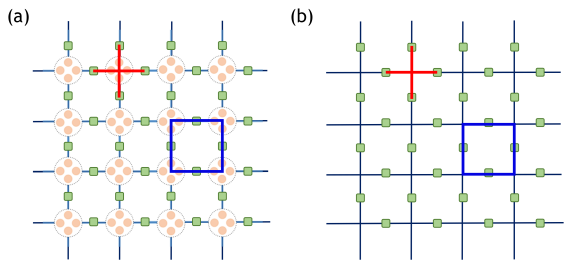}
\caption{Connection between the model in our studies to the
  $\mathbb{Z}_2$ gauge theory. The red star in (a) represents the
  expanded 4-body interacting term that is equivalent to the star term
  $\prod \sigma^{z}$ of the $\mathbb{Z}_2$ gauge model in (b).
  The blue plaquette in (a) represents the local combinatorial gauge
  generator $G_p$ by analogy with the local gauge generator
  $G_p^{\mathrm{toric}}$ of the $\mathbb{Z}_2$ gauge model defined as
  $\prod \sigma^{x}$ of the 4 spins around a plaquette in (b).}
\label{fig.mapping}
\end{figure} 

This local $\mathbb{Z}_2$ gauge symmetry is \emph{exact} for any
value of the parameters in the Hamiltonian
Eq.~\eqref{eq:Hgeneral}. We can obtain further intuition by connecting
to the more familiar formulation of the $\mathbb{Z}_2$ gauge
theory~\cite{wegner1971duality, RevModPhys.51.659} in certain
limits. Consider the effective Hamiltonian for the $H_s$ terms when
their energy scales are larger than those in $H^\sigma_{\rm kin}$; in
this regime, one can diagonalize $H_s$ by fixing the $\sigma_i^z$
around the star and treating the problem as that of a paramagnet for
the matter spins $\mu_a$. The result is an effective Hamiltonian for
the lowest states that take the form of a four-spin interaction among
the gauge spins:
\begin{subequations}
\begin{align}
  H^{\rm eff}_s = -\gamma -\lambda\; \prod_{i\in s} \sigma_i^{z}
  \;,
  \label{eq.lowEsec}
\end{align}
where the parameters $\gamma$ and $\lambda$ are given by~\cite{chamon2019emulate}
\begin{align}
  \gamma&=\frac{1}{2}\left(\sqrt{\Gamma_m^2 + 16J^2} + 3\Gamma_m + 4\sqrt{\Gamma_m^2 + 4J^2} \right) \\
  \lambda&=-\frac{1}{2}\left(\sqrt{\Gamma_m^2 + 16J^2} + 3\Gamma_m - 4\sqrt{\Gamma_m^2 + 4J^2}\right). \label{lambdadef}
\end{align}
\end{subequations}
Notice that the effective $H^{\rm eff}_s$ is, up to a constant shift,
the same as the star term $A_s^{z} = \prod_{i\in s}\sigma_i^{z}$ that
appears in the toric code~\cite{KITAEV20032} and the $\mathbb{Z}_2$
lattice gauge model~\cite{wegner1971duality, RevModPhys.51.659}. We depict in Fig.~\ref{fig.mapping}(a) the star
term in our model, juxtaposed to the star term represented in the
toric and $\mathbb{Z}_2$ gauge models in Fig.~\ref{fig.mapping}(b) as
the product of four spins on the red cross. The manifold of other
states in our model, those beyond the effective term, are separated by
a scale $\Gamma_m$. Thus, in the limit $\Gamma_m \to \infty$, higher
energy sectors are projected out, and the system Hamiltonian
asymptotically becomes the exact star term of the toric code.
	
In the conventional $\mathbb{Z}_2$ gauge transformation, the gauge
operator $G_p^{\mathrm{toric}} = \prod_{i \in p} \sigma_i^{x}$ is
built as a product of $\sigma_i^x$ spins along the smallest loops,
shown as the blue square in Fig.~\ref{fig.mapping}(b)). In a system
with Hamiltonian as in Eq.~\eqref{eq:Hgeneral}, which possesses
combinatorial $\mathbb{Z}_2$ gauge symmetry, the exact local gauge
transformation on a plaquette $p$ includes additional transformations
corresponding to the action of an operator $\mathcal{L}_s^{\mu}$ on
the matter spins of star $s$ as
\begin{align}
  \mathcal{L}_s^{\mu} \;\mu^z_a \;\left(\mathcal{L}_s^{\mu}\right)^{-1}
  =
  \sum_b \mu^z_a \;(\mathbf{L}^{-1})_{ba}
  \;,
\end{align}
which implements the flips and permutations associated to the monomial
matrix $\mathbf{L}$. The plaquette term generating the combinatorial
$\mathbb{Z}_2$ gauge symmetry is then defined as
\begin{align}
G_p = \prod_{s \in p}\mathcal{L}_s^{\mu}\prod_{i\in p}
\sigma_i^{x}.
\end{align} 
In a system with linear size $L$ (total spins
$N = L \times L \times 6$) and periodic boundary condition, one can
find $M = 2 + (L^2-1)$ independent gauge operators $G_p$ that commute
with the Hamiltonian Eq.~\eqref{eq:Hgeneral}. Within the $M$
operators, two of them, $G_x$ and $G_y$, are defined along
non-contractible loops in the two spatial directions, and their
quantum numbers uniquely characterize the topological ground state
degeneracy. Therefore in the basis of these $G_p$ operators
$G_p\ket{q} = q\ket{q}$ and the Hamiltonian can be block-diagonalized
into $2^M$ blocks associated with unique quantum number sets
$\mathbf{q} = (G_x=\pm1,G_y=\pm1;\pm1,\pm1,\dots)$ (See
Appendix.~\ref{app:Z2sym} for details), where we list the quantum
numbers of the non-contractible loops $G_x$ and $G_y$ first, and the
remaining quantum numbers are associate to the other $M-2$ independent
local gauge operators.

The discussion thus far is rather general, and, in particular, the
combinatorial gauge symmetry is exact, provided that the kinetic term
$H^\sigma_{\rm kin}$ involves only the $\sigma^x$ component. Below we
shall discuss two different choices of $H^\sigma_{\rm kin}$.

\subsection{Model with transverse field on the \\ gauge spins---model-X}
\label{sec:tfimgauge-model}

A simple choice of kinetic term is to apply a transverse field on
the gauge spins,
\begin{align}
  H^\sigma_{\rm kin} = 
  - \Gamma_g \sum_{i} \sigma^x_i
  \;,
  \label{eq:Hkin_TF}
\end{align}
or, equivalently, the case with full Hamiltonian
\begin{align}
  H =
  J\sum_{a\in s} \left(\sum_{j\in s} W_{aj} \sigma^z_j \right) \mu^z_a
  -\Gamma_m \sum_{a\in s} \mu^x_a
  -\Gamma_g \sum_{i} \sigma^x_i
  \;.
  \label{eq.WithTF}
\end{align}
\begin{figure*}[ht!]
\includegraphics[width=130mm]{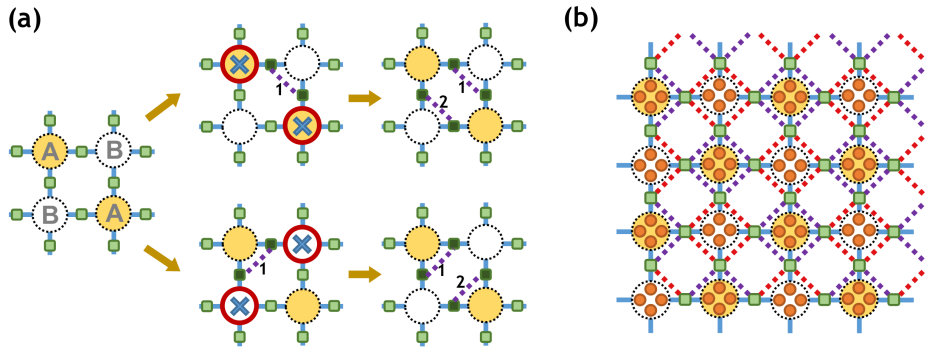}
\caption{The second-order process generates the effective plaquette term in model-XX. The A-sublattice is
  marked in light yellow while the B-sublattice is indicated in
  white. \textbf{(a)} The processes that couple two stars in the A
  (top part) or B (bottom part) sublattice with an effective
  ferromagnetic interaction. \textbf{(b)} The arrangement of XX bonds
  (dashed lines) resulting in an effective TFIM on A and B
  sublattices. Different colors indicate bonds coupling sites in the A
  and B sublattices.}
\label{fig.secproc}
\end{figure*}

\begin{figure*}[t]
\includegraphics[width=\textwidth]{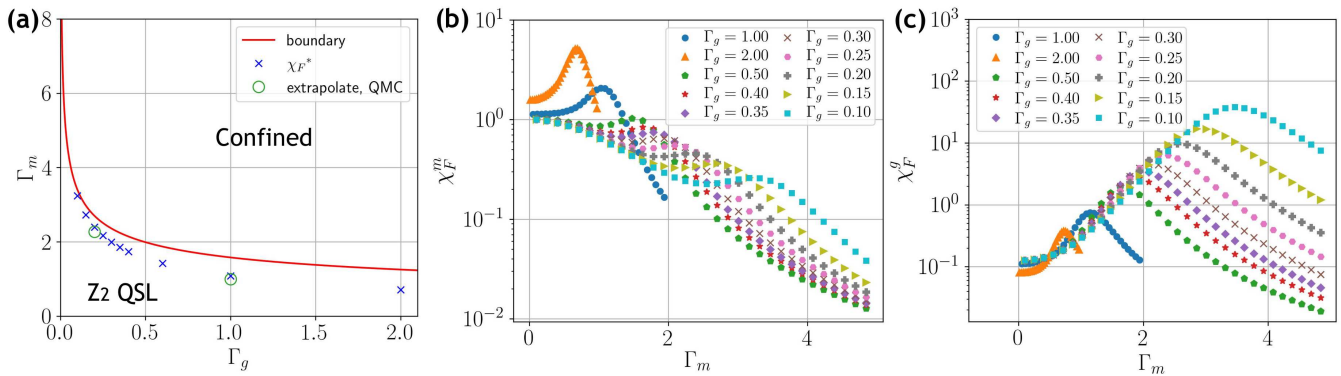}
\caption{(a) Phase diagram of model-X. The red curve is the perturbative large-$\Gamma_m$ phase boundary, Eq.~\eqref{gdef},
between $\mathbb{Z}_2$ quantum spin liquid and the confined (paramagnetic) phase, resulting from the mapping to the $\mathbb{Z}_2$
lattice gauge model. Blue crosses are the boundary points from the location of maximum fidelity susceptibility calculated with ED,
as shown in (b). The two open circles represent the $L\rightarrow \infty$ transition point extrapolated from the QMC data simulated
at fixed $\Gamma_g = 0.2$ and $1.0$, where the extrapolated transition points are at $\Gamma_m \approx 2.27$, and $\Gamma_m \approx 1.0$,
respectively, as discussed in Sec.~\ref{sec:topo1}. The Fidelity susceptibilities $\chi^{m}_F$ in (b) and $\chi^{g}_F$ in (c) were
calculated using Lanczos ED with $N = 2 \times 2 \times 6$ spins. }
\label{Fig.Xf}
\end{figure*}

In the limit $\Gamma_m \gg J$, we can replace the first two terms by
the star equivalent Eq.~\eqref{eq.lowEsec}. Therefore, Hamiltonian
Eq.~\eqref{eq.WithTF}, which obeys the exact local combinatorial
$\mathbb{Z}_2$ gauge symmetry, has the usual $\mathbb{Z}_2$ lattice
gauge model as its low energy description:
\begin{align}
\label{hz2gaugemodel}
H_{\mathbb{Z}_2}
=
-\lambda \sum_{s} \prod_{i \in s} \sigma^{z}_i
-g\sum_i \sigma^{x}_i,
\end{align}
with $g=\Gamma_g$ and $\lambda$ given by Eq.~(\ref{lambdadef}). The
$\mathbb{Z}_2$ lattice gauge model has been well studied and shown to
host $\mathbb{Z}_2$ topological order for $g < g_c$, and undergo a
deconfinement--confinement transition at $g=g_c$ that leads to a
paramagnetic phase for $g> g_c$. The critical point is
$g_c \approx 0.3285 \lambda$ from the exact mapping to the dual 2D
Ising model
\cite{Rieger1999,JonghDMRG1998,LiuQAQMC2013,fradkin_2013}. We expect
similar phases to appear in the full Hamiltonian Eq~\eqref{eq.WithTF}
as the uniform transverse field on the gauge spins is varied. The
limit $\Gamma_m \gg J$ therefore allows us to estimate the phase
boundary of our model Eq~\eqref{eq.WithTF} perturbatively by relating
the parameters $\Gamma_m, \Gamma_g$, and $J$ to the couplings $g$ and
$\lambda$ of the $\mathbb{Z}_2$ lattice gauge model using
Eq.~\eqref{eq.lowEsec}) and Eq.~\eqref{hz2gaugemodel} as
\begin{align}
\label{gdef}
\frac{g}{\lambda} \approx \frac{\Gamma_g}{12J^4/\Gamma_m^3}
=
\frac{\Gamma_g\Gamma_m^3}{12J^4}
.
\end{align}
Setting $g = g_c \approx 0.3285 \lambda$ results in the leading-order
phase boundary in the $(\Gamma_g,\Gamma_m)$ plane. By comparing with
unbiased numerical results, we will later show that this $\Gamma_m \to \infty$
form provides a good approximation to the phase boundary even down
to values as small as $\Gamma_m \approx 2$ [Fig.~\ref{Fig.Xf}(a)].

Note that the mapping to the simpler lattice gauge model is only
perturbatively exact in the limit of $\Gamma_m \rightarrow \infty$. In
the region with a small $\Gamma_m$, we cannot in general rule out other
phases appearing away from the perturbative regime. However, we do not
see any obvious reasons to expect phases beyond those of the effective
$\mathbb{Z}_2$ lattice gauge model. We present in
Sec.~\ref{sec:tfimgauge} our numerical results that support the
absence of other phases in our model and reveal the order of the
phase transitions between the topological and paramagnetic phases.

\subsection{Model with ferromagnetic XX-interaction on the gauge
spins---model-XX}
\label{sec:xxgauge-model}

Another simple choice of kinetic term is to add two-spin $XX$ interactions
between nearest-neighbor gauge spins,
\begin{align}
  H^\sigma_{\rm kin} = 
  - J_x \sum_{\langle kl\rangle} \sigma^{x}_k\sigma^{x}_l
  \;,
  \label{eq:Hkin_XX}
\end{align}
so that the full Hamiltonian is
\begin{align}
  H =
  J\sum_{a\in s} \left(\sum_{j\in s} W_{aj} \sigma^z_j \right) \mu^z_a
  -\Gamma_m \sum_{a\in s} \mu^x_a
  - J_x \sum_{\langle kl\rangle} \sigma^{x}_k\sigma^{x}_l
  \;.
  \label{eq.WithXX}
\end{align}
Notice that $H^\sigma_{\rm kin}$ satisfies the general conditions
presented above to retain the combinatorial gauge
symmetry. ($H^\sigma_{\rm kin}$ is written in terms of $\sigma^{x}$
only, and thus commute with the local operators $G_p$.)

To gain intuition about the model with Hamiltonian
Eq.~\eqref{eq.WithXX}, we again consider the limit $\Gamma_m \gg J$ in
which the first two terms can be replaced by their equivalent star
term Eq.~\eqref{eq.lowEsec}. Defining the four-spin star operator
$A^{z}_s\equiv \prod_{i \in s} \sigma^{z}_i$, the effective
Hamiltonian, without the kinetic term, reads
$H^{\rm eff}_0 = -\lambda \sum_s A^{z}_s$. The ground states of
$H^{\rm eff}$ have $A^{z}_s = 1$ (parity $+1$) on all stars or
vertices. Similarly to the dual mapping of the conventional
$\mathbb{Z}_2$ lattice gauge model, we can introduce a conjugate star
operator $A^{x}_s $ that flips the eigenvalue of $A^{z}_s$, in terms
of which we write the gauge spin between two nearest-neighbor stars
$s$ and $s'$ as $\sigma_i^{x} = A_s^{x}A_{s'}^{x}$.

The $XX$ kinetic term Eq.~\eqref{eq:Hkin_XX} perturbatively generates
four-spin plaquette interactions (i.e., products of four $\sigma^x$
around the small loop around a plaquette). To this end, the bond
operators in the kinetic term can be arranged, e.g., in the way
illustrated in Fig.~\ref{fig.secproc}. Starting from the ground state
of $H^{\rm eff}_0$, where all vertices have parity $A^{z}_s=+1$,
acting with a term of the form $\sigma^{x}_k\sigma^{x}_l$ on two sites
$k$ and $l$ within a vertex $s$ generates a pair of defects on either
the A or the B sublattice, depending on the bond chosen. Adding
another bond operator within the same plaquette, parallel to the first
bond, generates a plaquette term at second order in $J_x/\lambda$. The
effective perturbative Hamiltonian in terms of the $A^{z,x}$ operators
then takes the form
\begin{align}
  H^{\rm eff}
  =
  - \lambda \sum_s  A^{z}_s 
  - 2J_x \sum_{\langle s s'\rangle}
  A^{x}_s \,A^{x}_{s'}
  \;,
\end{align}
where $\langle ss'\rangle$ indicates nearest-neighbor stars within the
same sublattice, and $\lambda$ is the aforementioned
effective strength of the star-term, Eq.~\eqref{lambdadef}. In the
limit $\Gamma_m \gg J$, our model effectively reduces to two
independent TFIMs; one on each sublattice.  The particular arrangement
of the $XX$ terms imposes an additional even-odd conservation law in
the system. The parity of negative $A^{z}_s$ vertices within
a sublattice is conserved, which can be easily seen from
Fig.~\ref{fig.secproc}(a), where the $XX$ operators can only create defects in
pairs within each one of the sublattices.

In the limit $J_x \rightarrow \infty$, since all gauge spins
interact ferromagnetically, an $x$-direction ferromagnetic phase
arises; thus we expect a quantum phase transition between the
topological phase and a ferromagnetic phase, replacing the
topological--paramagnetic transition of the model-X discussed in
Sec.~\ref{sec:tfimgauge-model}. Similarly to the model with transverse
field on the gauge spins, where the phase boundary between the
topological and paramagnetic phases is given perturbatively by
Eq.~\eqref{gdef} through the mapping to the TFIM, here this mapping
gives the following relation between the field strength in the
TFIM and the field $\lambda$ in Eq.~\eqref{lambdadef};
\begin{align}
\frac{\lambda}{2J_x} \approx \frac{6J^4}{\Gamma_m^3J_x}.
\end{align}
Thus, the topological--ferromagnetic phase boundary can be obtained to
leading order by setting the ratio above to the critical point $3.04497(18)$ of the 2D
quantum Ising model~\cite{LiuQAQMC2013,JonghDMRG1998}. This boundary also
describes numerical results for surprisingly small values of the matter
field, down to $\gamma_m \approx 1$ [Fig.~\ref{Fig.Wl.Xf}(a)].

\section{Analysis of model-X}
\label{sec:tfimgauge}

In this section, we present our numerical studies of the model-X
introduced in Sec.~\ref{sec:tfimgauge-model}. Our results support the
theoretical conjecture that the model has a topological and a
paramagnetic phase with no other phases. The nature of the quantum
phase transitions between these two states is revealed. In the following, we
fix $J=1$ and impose periodic boundary conditions in all our numerical
simulations.

\subsection{Fidelity susceptibility}

To confirm that the model does have phases predicted by the effective
$\mathbb{Z}_2$ gauge theory, we start by identifying signatures of the
phase transition. We first consider the fidelity susceptibility
 which can probe the existence of a phase
transition without requiring knowledge of any order parameter~\cite{YouFidelity2007}. The
fidelity susceptibility is defined as the second derivative of the
logarithmic fidelity with respect to a generic tuning parameter $x$
\begin{align}
\chi_F = \frac{\partial^2\ln F_x}{\partial \delta_x^2}\vert_{\delta_x=0},
\end{align}
where $F_x = |\left< \psi(x)|\psi(x+\delta_x)\right>|$ is the infinitesimal fidelity in the direction defined by $x$.  In our model with transverse fields,
two types of fidelity susceptibilities can be formally defined with variations along the two different transverse fields;
$x=\Gamma_m$ or $x=\Gamma_g$;
\begin{subequations}
\begin{align}
  \chi^{m}_F(\Gamma_m,\Gamma_g) &= \frac{\partial^2\ln F_{\Gamma_m}}{\partial \delta_{\Gamma_m}^2}\vert_{\delta_{\Gamma_m}=0}, \\ 
  \chi^{g}_F(\Gamma_m,\Gamma_g) &= \frac{\partial^2\ln F_{\Gamma_g}}{\partial \delta_{\Gamma_g}^2}\vert_{\delta_{\Gamma_g}=0}. \label{eq.Xf2}
\end{align} 
\end{subequations}
Across a phase transition, the fidelity susceptibility should develop a maximum that diverges in the thermodynamic limit. 

We first calculate the fidelity susceptibilities using exact diagonalization with the Lanczos method \cite{GUXf2010,PrelovsekLanczos2013} on a small system
with $2\times 2$ unit cells, i.e., $N=24$ spins in total. Due to the rapid growth of the Hilbert space, this is currently the largest accessible system size
with our computational resources. Fig.~\ref{Fig.Xf}(b) shows $\chi^{m}_F$ as a function of $\Gamma_m$ for several different values of $\Gamma_g$.
A single peak is present in all cases, which implies the possibility of a phase transition in the thermodynamic limit. Furthermore, we have not
observed any cases of multiple maxima in any of our calculations, suggesting only two different phases. Similar behaviors are also observed for
$\chi^{g}_F$ as shown in Fig.~\ref{Fig.Xf}(c). From the location of the maximum of $\chi^{m}_F$ shown in Fig.~\ref{Fig.Xf}(a), we find that the data
fall close to the perturbative (large-$\Gamma_m$) topological--paramagnetic phase boundary even though the value of $\Gamma_m$ is not extremely large (and
$\Gamma_g$ not extremely small).

It may seem surprising that the phase boundary is given accurately by a system with only four unit cells. To confirm that the observed maximum 
of the fidelity susceptibility grows with the system size and truly indicates a phase transition, we next turn to QMC simulations to reach 
larger system sizes. We use the SSE QMC method \cite{Sandvik2010,s2019stochastic}, for which a convenient way to compute the fidelity susceptibility
was devised recently \cite{Lei:2015}.

\begin{figure}[t]
\includegraphics[width=75mm]{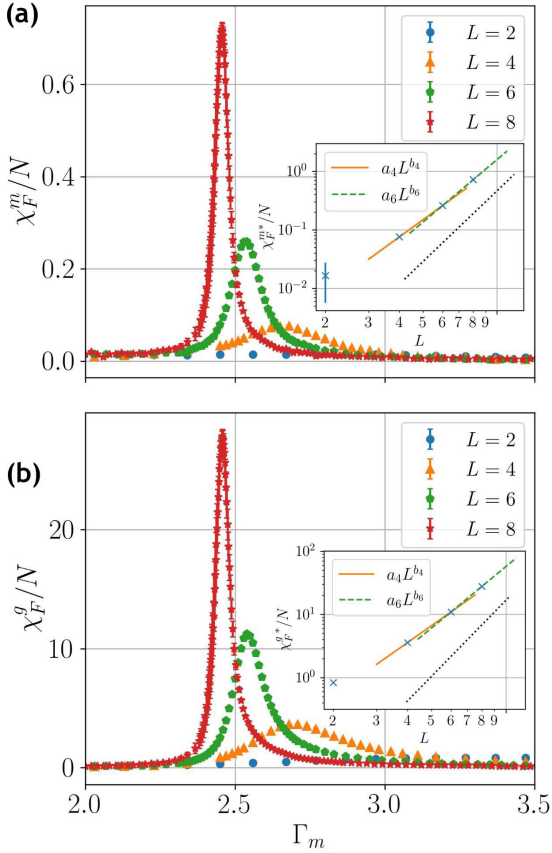}
\caption{SSE results for the size-normalized fidelity susceptibilities $\chi^{m}_F/N$ (a) and $\chi^{g}_F/N$ (b) for systems with different number
of spins $N = L \times L \times 6$. All data points for a given $L$ were obtained in the same simulation with quantum parallel tempering at fixed
$\Gamma_g = 0.2$ and inverse temperature $\beta = 4L$. The insets show log-log plots of the peak value of the fidelity susceptibility versus $L$, 
along with solid lines drawn through pairs of data points with $L$ and $L+2$ to analyze power-law behaviors $\chi_F/N \sim L^{b_L}$. 
The extracted size-dependent exponents in (a) are $b_4\approx 3.0$ from the $L=4,6$ points and $b_6\approx 3.5$ from $L=6,8$. In (b) the $L=4,6$ points give 
$b_4 \approx 2.7$ and $L=6,8$ give $b_6 \approx 3.2$. 
The dotted lines in the insets of (a) and (b) have a slope $b=2(d+1)-d=4$ corresponding to a first-order transition and are shown as a reference.}
\label{Fig.Wl}
\end{figure}
	    
In our simulation, we set the inverse temperature as $\beta = 4L$. Because of the small vison gaps in the topological phase, this scaling of $\beta$ does 
not allow us to reach the finite-size ground state deep inside the topological phase. However, with the $T \to 0$ limit approached with $\beta \propto L$
we can still address the nature of the quantum phase transition from the gapped paramagnetic phase. In our model, the Ising interactions are highly frustrated, 
and to mitigate the associated effects of slow dynamic of the QMC updates in the topological phase and at the phase transition, we have implemented 
quantum replica exchange~\cite{Hukushima:1996RX,Hukushima:1996RXinIJ,Sengupta:2002QRX}. Simulations are thus carried out in parallel for a large number of replicas with different values of $\Gamma_m$ 
on both sides of the transition, 
with swap attempts carried out for neighboring values of the 
parameter after several conventional SSE updates. Even with replica exchange, it is still difficult to equilibrate systems for large $L$, and we have limited 
the present study to $L \le 8$. As we will see, these moderate system sizes are already sufficient for drawing definite conclusions.

Fig.~\ref{Fig.Wl}(a) shows the results of the fidelity susceptibility $\chi^{m}_F$ at $\Gamma_g = 0.2$ as a function of $\Gamma_m$. We indeed find that the
peak identified in the ED calculations diverges upon increasing the system size, providing solid evidence of a phase transition. 
The other fidelity susceptibility $\chi^g_F$ shows a similar behavior, as shown in Fig.~\ref{Fig.Wl}(b).

In order to understand the nature of the phase transition, we perform finite-size scaling of the maximum value of the fidelity susceptibility to extract
the associated critical exponent. Based on the similarity of the model to the $\mathbb{Z}_2$ lattice gauge model, in which the transition is in the (2+1)D Ising 
universality class, one might naively expect a continuous transition at which the maximum should scale with system size as~\cite{GUXf2010} 
$\chi_F/L^{d} \sim L^{2/\nu - d}$ with $d=2$ the spatial dimensionality and $\nu \approx 0.63$. 
However, we do not observe a scaling of the above form. Instead, we
analyze the data using a generic scaling form
$\chi_F/L^{d} \sim L^{b}$ with an adjustable exponent $b$. To further
take into account finite-size corrections, we consider a size
dependent exponent $b_L$ extracted from two system sizes, $L$ and
$L+2$; graphically this exponent corresponds to the slope of the line
drawn between two data points on a log-log scale as shown in the
insets of Fig.~\ref{Fig.Wl}.

In the case of $\chi_F^m$ in Fig.~\ref{Fig.Wl}(a) we find
$b_4 \approx 3.0$ (i.e., the line drawn between data points for $L=4$
and $L=6$) and $b_6 \approx 3.5$ (from $L=6,8$). In the case of
$\chi^g_F$ we find $b_4 \approx 2.7$ and $b_6 \approx 3.2$. These
exponents are significantly larger than the expected value
$2/\nu - d\approx 2/0.63-2 \approx 1.175$ of the (2+1)D Ising
universality class, and for both susceptibilities the deviation
becomes larger for the larger system sizes.  It therefore appears more
likely that the transition is first-order. Generally, at classical
first-order transitions the same scaling forms hold as for continuous
transitions, but with the exponent $1/\nu$ replaced by the
dimensionality $d$ \cite{FisherFOscal,BinderLandauFO}. In a quantum
system, the replacement should be $1/\nu \rightarrow d+z$, where the
appropriate value of the dynamical exponent $z$ reflects the nature of
the low-energy excitations in the two coexisting phases
\cite{Anders2010FOQT,Zhao2019}. Our results in Fig.~\ref{Fig.Wl}
suggest a first-order behavior with $z=1$, in which case
$b = 2/\nu - d \rightarrow 2(d+1)-d = 4$. We show this type of
divergence for reference with the dotted lines in the insets of
Figs.~\ref{Fig.Wl}(a) and \ref{Fig.Wl}(b); this asymptotic behavior
seems very plausible based on the available data.

\subsection{Topological order}
\label{sec:topo1}

\begin{figure}[t]
\includegraphics[width=75mm]{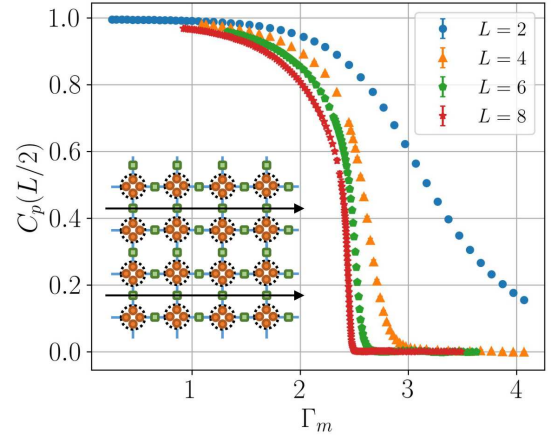}
\caption{Wilson-loop correlation function obtained in SSE simulations with $\Gamma_g = 0.2$. Data for different system sizes show
convergence to a finite value for $\Gamma_m \agt 2.3$, with the expected value $C_p(L/2)=1$ in the $\Gamma_m \to 0$ limit. The inset illustrates 
the definition Eq.~\eqref{eq.Cp} of the correlation function in terms of two parallel non-contractible Wilson loops in the $x$ direction
of a periodic lattice of size $L=4$.}
\label{Fig.Wlb}
\end{figure}

Next, we turn to the properties of the underlying phases. Based on the
mapping to the $\mathbb{Z}_2$ lattice gauge model, we expect the phase
with small $\Gamma_m$ to be a $\mathbb{Z}_2$ topological quantum spin
liquid. Note that Elitzur's theorem forbids any spontaneous symmetry
breaking of local gauge symmetries; thus one cannot define any local
order parameter to characterize such topological
order~\cite{Elitzur1975,Kotecky1992,RevModPhys.51.659}. To detect the
topological order, we investigate the global, non-contractible Wilson
loop operator, defined as the product of gauge spin $\sigma^z$
operators along a non-contractible loop in the $\alpha$-direction
\begin{align}
p_{\alpha,n}= \prod_{\{i\}_{\alpha,n}} \sigma_i^z,~~~\alpha \in \{x,y\},
\end{align}
for the set of sites $\{i\}_{\alpha,n}$ belonging to the $n$-th row or column.
For a $\mathbb{Z}_2$ spin liquid, the quantum numbers
$p_{x,n} = \pm 1$, $p_{y,n} = \pm 1$ characterize the four degenerate
(in the thermodynamic limit) topological ground states regardless of
which row or column $n$ is chosen. We can take advantage of this
property to define a correlation function detecting the topological
order using the product of two parallel non-contractible loops on rows
or columns labeled by $m$ and $n$:
\begin{align}
c_p^{\alpha}(r_{mn})=\langle p_{\alpha,m}p_{\alpha,n}\rangle = \left(\prod_{\{i\}_{\alpha,m}}\sigma_i^z\right) \left(\prod_{\{i\}_{\alpha,n}}\sigma_i^z\right),
\end{align}
which we also illustrate in the inset of Fig.~\ref{Fig.Wlb}. Instead of investigating this correlation as a function of the distance $r_{mn}$ between the two
loops, we here take the longest distance for a given lattice size, $r_{mn}=L/2$, and analyze the $L$ dependence of $C_p(L/2)$ defined as a summation
over all translations (to reduce the statistical fluctuations) of the two Wilson loops oriented in the $\alpha \in \{ x,y\}$ lattice direction: 
\begin{align}
C_p(L/2) \equiv \frac{1}{L} \sum_{j=1}^L c_p^{\alpha}(r_{j,j+L/2}), \label{eq.Cp}
\end{align}
which can be averaged over the two directions. In the topologically ordered phase we expect $C_p(L/2) \neq 0 $ when $L \to \infty$, while in the
paramagnetic phase $C_p(L/2)\to 0$. Note that this quantity has been used in a previous study of topological order in classical Ising gauge models
at zero and non-zero temperatures \cite{Na:2018}. 

In Fig.~\ref{Fig.Wlb}, we show SSE results at $\Gamma_g = 0.2$ as a function of $\Gamma_m$. We see that $C_p(L/2)$ indeed vanishes with increasing $L$ for large 
$\Gamma_m$, while it converges to a finite value for $\Gamma_m$ in a range consistent with the transition point found above for the same value of $\Gamma_g$. 
Below we will discuss the size-extrapolated phase boundary. We stress here that the Wilson loop order parameter does not detect any phases with only local 
order parameters (See Appendix.~\ref{app:2DTFIM} for a study of the ferromagnetic state as an example) and our results therefore demonstrate conclusively
a $\mathbb{Z}_2$ topological phase of finite extent as the field $\Gamma_m$ is turned on.

\begin{figure}[t]
\includegraphics[width=75mm]{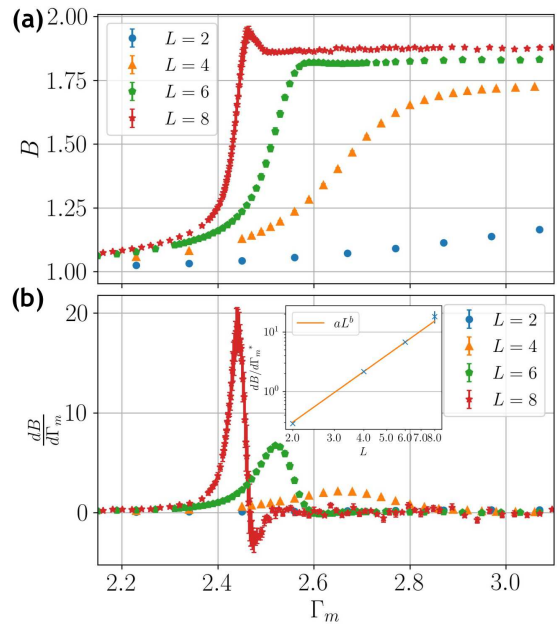}
\caption{Results for the model with transverse field at $\Gamma_g = 0.2$. (a) The Binder ratio $B$ defined with both components of the 
Wilson loop order parameter $P^2 = P_x^2 + P_y^2$ in Eq.~\eqref{bratiodef_xy}. The peaks adjacent to the phase transition for system sizes $L \geq 6$
(barely discernible for $L=6$)
are signatures of a first-order transition. (b) The numerical derivative of $B$ with respect to $\Gamma_m$  [computed using the linear approximation 
between the successive points in (a)]. The inset shows a power-law fit $y \propto L^{b}$ to the maximum value of the derivative, with only the largest 
three system sizes included. The exponent is $b \approx 3.0$, which is consistent with expected value $b=d+1 = 3$ for a first-order transition.}
\label{fig.B2}
\end{figure}

Having established a good topological order parameter, we further provide evidence of a first-order phase transition by a finite-size scaling analysis of
the corresponding Binder ratio. To this end, we define the topological order parameter on the entire system as the sum of Wilson loops with
$P^2 = P_x^2 + P_y^2$ where
\begin{align}
P_\alpha = \frac{1}{L}\sum_{n=1}^{L} p_{\alpha,n},
\label{pxydef}
\end{align}
with $\alpha \in \{x,y\}$,
and the Binder ratio
\begin{align}
B = \frac{\langle P^4\rangle}{\langle P^{2}\rangle^2}. 
\label{bratiodef_xy}
\end{align}  
In a perfect $\mathbb{Z}_2$ ordered topological phase, $P_x = \pm 1$
and $P_y = \pm 1$, forming a $\mathbb{Z}_2\times \mathbb{Z}_2$
symmetric order parameter distribution, while $P_x=P_y=0$ in the
paramagnetic phase. With increasing system sizes, the Binder ratio is
expected to form a step function at the transition point in the case
of a continuous transition, while the distribution of the order
parameter in the coexistence state at a first-order transition
typically is also associated with a divergent peak adjacent to the
step \cite{Vollmayr1993:FOT,Iino:FOPT}. In Fig.~\ref{fig.B2}(a) the
Binder ratio indeed evolves into a step function with a side peak,
though the latter is only seen clearly for the largest system sizes,
$L=8$, and for $L=6$ there is a very weak maximum as well. Looking at
the derivative of $B$, in Fig.~\ref{fig.B2}(b) we observe a divergent
positive main peak, and for $L=8$ a prominent negative peak reflects
the presence of the first-order side peak in
Fig.~\ref{fig.B2}(a). Thus, we have strong evidence of phase
coexistence at a first-order quantum phase transition caused by an
avoided level crossing. In Appendix~\ref{app:TFpxpy}, we further show
the full distribution of the two-component Wilson loop order parameter
$\rho(P_x, P_y)$, which clearly shows a phase coexistence
characteristic of a first-order transition.

The maximum of the Binder ratio derivative diverges with the system size as the step function develops. The derivatives can be evaluated directly in the SSE 
simulations, using an estimator derived in Appendix~\ref{app:dB2}. With the replacement $1/\nu \to d+z$ in the scaling form
$dB/d\Gamma_m \sim L^{1/\nu}$ and expecting $z=1$ here, the peaks should diverge as $L^3$. Indeed, in Fig.~\ref{fig.B2}(b) the peak values for the three largest system sizes can be fitted to a power-law $L^{b}$ with $b \approx 3.0$, thus supporting a first-order
transition. 

\begin{figure}[t]
\includegraphics[width=75mm]{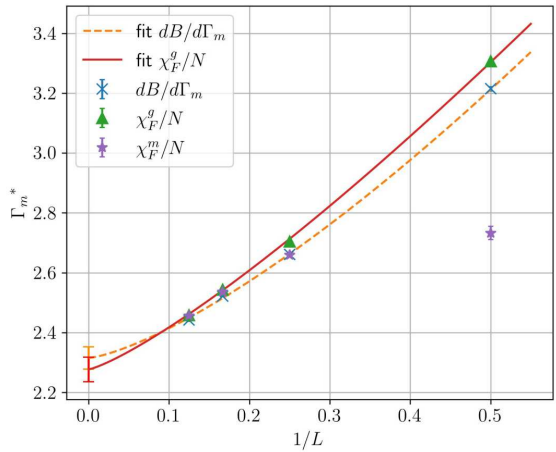}
\caption{Scaling with the inverse system size of the $\Gamma_m$ values of the finite-size maxima $dB/d\Gamma_m$, $\chi^{m}_F/N$ and $\chi^{g}_F/N$,
with $\Gamma_g = 0.2$. Both $dB/d\Gamma_m^*$ and $\chi^{g*}_F/N$ have been fitted with a single power-law correction
and give the $L\to \infty$  extrapolated values $\Gamma_{m}^{c} = 2.31(3)$ and $\Gamma_{m}^{c}= 2.27(4)$, respectively. The apparent large subleading 
corrections to $\chi^{m*}_F$ location forbid us to get a reasonable extrapolation based on the available data, though the $L=6$ and $L=8$ pints show full 
consistency with the other estimates.}
\label{fig.extra}
\end{figure}

So far, we have discussed the divergence properties of the peaks in
the Binder ratio and the fidelity susceptibility. We also need to
extrapolate the peak locations in order to obtain the transition point
in the thermodynamic limit. Fig.~\ref{fig.extra} shows the dependence
of the peak locations on $1/L$ along with extrapolations assuming
power-law corrections. All quantities show mutually consistent
behaviors for the largest system sizes, but $\chi_F^m$ has much larger
scaling corrections than the other quantities. Extrapolations with
error analysis give the critical value of the matter field
$\Gamma_{m}^{c} \approx 2.27$ for the gauge field $\Gamma_g=0.2$
considered here. In the phase diagram in Fig.~\ref{Fig.Xf}(a) we have
marked this transition point with a circle, and we also show the result
obtained using the same methods for $\Gamma_g=1$. These QMC points are
very close to the boundary estimated from the ED results for a very
small system with $L=2$.

Here we should note that the ED results are
calculated exactly at $T=0$, while there are still some temperature
effects left in the QMC results obtained with our choice of temperature
scaling, $T=(4L)^{-1}$. In the
case of $L=2$, the QMC results for $\Gamma_{m}^{c}(L)$ are actually
quite far from the $T=0$ ED results because of the temperature
effects. However, since $T \to 0$ as $L$ increases, the $L \to \infty$
extrapolated QMC results are not affected by finite temperature (beside
unimportant constant factors in the peaks of the physical quantities).
In this regard, it can also be noted that effects of inappropriate
temperature scaling with $L$ could potentially ruin a quantum phase
transition that does not extend to $T>0$ (as is the case with
topological order in two spatial dimensions), while there is no reason
to expect a transition detected when $T \propto 1/L$ to vanish if $T$
approaches zero more rapidly.

\subsection{Energy derivative}

\begin{figure}[t]
\includegraphics[width=75mm]{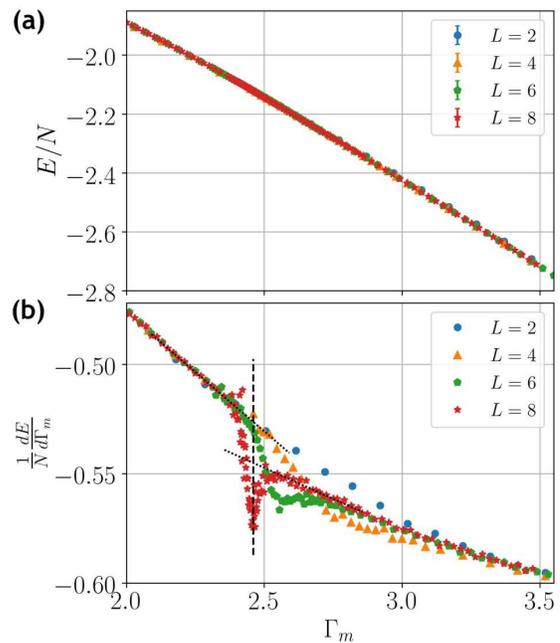}
\caption{(a) Energy density computed in SSE simulations with $\Gamma_m = 0.2$. 
(b) The corresponding derivatives are taken numerically based on the available data in (a).
Features indicating a discontinuity developing with increasing $L$ demonstrate
a first-order transition. We have fitted lines to the $L=8$ data away from the
sharp features and observe the presence of a jump when these forms are extrapolated
to the location of the sharp peak (the vertical dashed line, which can be taken as a
finite-size definition of the transition point).}
\label{FigdEtfim}
\end{figure}

In this section, we show another signature of the first-order phase
transition from the ground state energy density. In practice, the
$T \to 0$ internal energy is obtained with QMC calculations at
$T=(4L)^{-1}$.  The type of first-order transition indicated by the
results above, where finite-size scaling with the exponent replacement
$1/\nu \to d+z$ holds, should be associated with an avoided level
crossing. Thus we expect a change in the slope of the energy at the
transition with increasing $L$. In Fig.~\ref{FigdEtfim}(a) we plot the
energy per site as a function of $\Gamma_m$ for the same gauge-field
strength as considered above, $\Gamma_g = 0.2$. At first sight, the
data exhibit a smooth behavior without any visible kink. However, by
taking the numerical derivative of the energy, as shown in
Fig.~\ref{FigdEtfim}(b), we find a clear signature of non-analytic
behavior, such that the derivative becomes discontinuous at the
transition in the thermodynamic limit.

Along with the other results, this demonstration of a discontinuous
energy derivative provides definite proof of a first-order quantum phase
transition between the topological and paramagnetic states.
	
\subsection{Level spectroscopy}

Having used QMC simulations to establish the existence of an extended $\mathbb{Z}_2$ topological phase and its quantum phase transition into the
paramagnetic phase, we now again turn to Lanczos ED calculations in order to investigate the energy level spectrum of the system. We use the
combinatorial $\mathbb{Z}_2$ symmetry to block-diagonalize the Hamiltonian into $M = 2+L^2-1$ blocks in the basis of $\mathbb{Z}_2$ gauge generators,
as discussed in detail in Appendix~\ref{app:Z2sym}. The blocks are categorized by a set $\mathbf{q} = (\pm,\pm,\{\pm\})$ of quantum numbers, where the
first two elements correspond to the two non-contractible loops with associated quantum numbers $G_x$ and $G_y$ and $\{\pm\}$ denotes the set of
$L^2-1$ local quantum numbers $G_i$. In the thermodynamic limit, the topological ground state should be four-fold degenerate, corresponding to the
lowest energy states from sectors with $G_x = \pm$, $G_y = \pm$ and $G_i = +$ for all other $i \in M-2$ local operators. The transverse field does
not commute with the Hamiltonian; thus, there are always finite-size gaps between the four topological states in a finite system. Our ED calculations
here are again restricted to $L=2$ (and we present some QMC results also for $L=4$), but even for this very small system many of the salient
signatures of spinon and vison excitations can be observed, as well as signatures of the first-order quantum phase transition.

\begin{figure}[t]
\includegraphics[width=75mm]{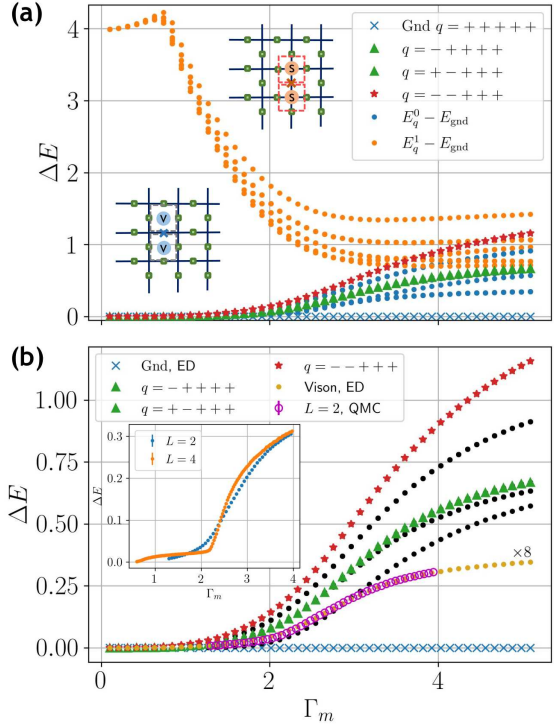}
\caption{ED level spectrum at $\Gamma_g = 0.2$ for a system of size $L = 2$ ($N = 2 \times 2 \times 6$ spins). In (a) the two lowest energy gaps
relative to the ground state (marked Gnd in the legends) are graphed versus the matter field strength for each of the 32 blocks with quantum numbers
$\mathbf{q} = (\pm,\pm,\pm,\pm,\pm)$. Many blocks are degenerate because of lattice symmetries; thus the number of different curves is much less than
64. The four states that become degenerate topological ground states in the $\mathbb{Z}_2$ phase are marked by blue crosses (the finite-size ground
state), red stars, and green triangles (two degenerate sectors); these states all have the local quantum numbers $G_i=+1$. The blue dots represent all
other lowest block levels; these are states with visons (two or a larger even number) of the topological phase (marked by $v$ in the inset illustration).
The orange curves represent the second-lowest states in each block; they correspond to the spinon excitations (particles indicated in the inset by $s$)
of the topological phase and they all become degenerate for $\Gamma_m \to 0$. In (b), the lowest block states are graphed on a magnified scale. The
eight-fold degenerate level that is the lowest excitation in the paramagnetic phase was calculated with both ED (yellow dots) and extracted from
imaginary-time correlations from QMC simulations (violet circles), to demonstrate the correctness of the latter for $L=2$. QMC results for both $L=2$
and $L=4$ (calculated at inverse temperature $\beta=24L$)
are shown in the inset.}
\label{Fig.Vgap_tfim}
\end{figure}
		
In Fig.~\ref{Fig.Vgap_tfim}(a), we graph low-energy gaps $\Delta_E$ relative
to the ground state versus the matter field $\Gamma_m$ at fixed
$\Gamma_g = 0.2$. For each of the 32 topological  symmetry blocks
pf the $L=2$ system, the two smallest gaps are shown, but because of
degeneracies due to lattice symmetries there are only 11 distinct curves.
The unique finite-size ground state has $\mathbf{q} = (+,+,+,+,+)$, i.e.,
$G_x=+$, $G_y=+$ and $G_i=+$ for $i=1,2,3$. The four levels that
become degenerate in the thermodynamic limit in the topological phase
are highlighted with different symbols. As for the remaining
low-energy levels, we note that in the $\mathbb{Z}_2$ topological
phase two types of quasi-particle excitations should be expected;
spinons ($s$) and visons ($v$), which are created in pairs by acting
on the ground state with $\sigma^x$ and $\sigma^z$ respectively on the
gauge spins, as indicated in the insets of
Fig.~\ref{Fig.Vgap_tfim}(a).

In the $\Gamma_m \rightarrow 0$ limit, the spinon excitations are gapped with $\Delta_E \approx 4J$ (the gap value is $4J$ if $\Gamma_g = 0$), as seen clearly
in Fig.~\ref{Fig.Vgap_tfim}(a), where these levels are shown with orange symbols. The vison gap opens when increasing $\Gamma_m$, as can be
seen from the fact that the effective model takes the form
\begin{align}
H_{pert} = \sum_s H_s - g\sum_p G_p
\label{hpert}
\end{align}
to lowest order in perturbation theory. Here $G_p$ is the local gauge generator that appears at 12th order, where the coupling is
\begin{align}
g \propto \left(\Gamma_m^8 \Gamma_g^4\right)/J^{11}.
\label{ghpert}
\end{align}
We can identify the vison excited states simply by considering the quantum number blocks that
couple to the ground state through the on-site $\sigma^z$ operators (which do not commute with the local gauge operators). In Fig.~\ref{Fig.Vgap_tfim}(b),
we plot ED results for the same parameters as in Fig.~\ref{Fig.Vgap_tfim}(a), but with a change in scale to focus on the vison states. These states
are gapped for all $\Gamma_m > 0$, but the gaps are much smaller in the topological phase than in the paramagnetic phase. The lowest vison state, which
contains two visons, is eight-fold degenerate on the small system considered here. The other levels in Fig.~\ref{Fig.Vgap_tfim}(b) correspond to
states with (an even number) more than two visons.

We can extract the lowest vison gap from QMC simulation by analyzing the imaginary-time autocorrelation function of $\sigma_z$, defined as
\begin{align}
G(\tau) = \frac{1}{N_g} \sum_{i \in g} \left< \sigma^z_{i}(0)\sigma^z_{i}(\tau) \right>,
\end{align}
where $N_g = L \times L \times 2$ is the total number of gauge spins in the system. The gap can be extracted by fitting an exponential
function to $G(\tau)$ for large $\tau$. Here it is also important that the temperature is sufficiently low, so that the asymptotic form of $G(\tau)$ 
is dominated by the lowest gap; See Appendix.~\ref{app:gapXX} for further technical details on these calculations. In Fig.~\ref{Fig.Vgap_tfim}(b) we 
compare the lowest gap extracted from the QMC data for the $L=2$ system at inverse temperature $\beta=48$ with the ED result. We observe good agreement 
between the two calculations. Note that the eight-fold degenerate levels with two vison excitations
undergo a true (not avoided) level crossing with a state with $G_x=G_y=+$ and all local quantum numbers $G_i=-$. 
This level-crossing is a finite-size effect, and we do not expect such behavior to persist for larger system sizes.
The $G_i=-$ state contains four visons, i.e., it can be reached from the ground state with two different 
$\sigma^z_i$ operations. It therefore does not contaminate the correlation function $G(\tau)$ corresponding to the
two-vison level of interest. For this small system the four-vison state falls under the lowest 2-vison state below $\Gamma_m \approx 3$,
i.e., close to the phase transition into the topological state.

Note again that the quantum numbers $G_i$ are conserved (i.e. commute with the Hamiltonian) in both the topological phase and the paramagnetic phase
of the model. However, visons with $G_i=-$ are deconfined only in the topological phase. In the paramagnetic phase, the lowest energy vison excitations
must form bound states residing on two adjacent plaquettes, while states with more visons and larger separations between the visons have larger energy
costs, as shown in Fig.~\ref{Fig.Vgap_tfim}(b).

In the inset of Fig.~\ref{Fig.Vgap_tfim}(b), we show the vison gap based on QMC calculations for both $L=2$ and $L=4$ at $\Gamma_m=0.2$, using
inverse temperature $\beta = 24L$. While the $L=2$ gap exhibits only a rather smooth variation with $\Gamma_m$, at $L=4$ a sharp feature has developed
close to the phase transition. The sharp behavior of the gap here is consistent with the scenario of a first-order transition through an avoided
ground state level crossing, and this mechanism should be associated also with avoided level crossings of the low-lying excitations.

Although the model possess $\mathbb{Z}_2$ topological order and can be directly implemented on existing quantum devices \cite{zhou2020building},
it will be difficult to reach the true ground state, or even a thermal state with a low density of visons, due to the very small vison gap.
These difficulties are clear from the effective model obtained perturbatively, Eq.~\eqref{hpert} with the 12th-order effective coupling
in Eq.~\eqref{ghpert}.  Nonetheless, there may still be signatures of the mutual statistics of
the spinons and visons that could be observed in the regime where
temperature is larger than the vison gap but still much smaller than
the spinon gap, as discussed in Ref.~\cite{Hart2021}. In this regime
the visons randomly appear within plaquettes because their energy of
formation is much smaller than the temperature. In the presence of kinetic
terms (such as a transverse field), the spinons acquire dynamics at a
scale much faster than that of the visons, so effectively they quantum
diffuse in a background or randomly placed visons. Because of the
mutual statistical phase of $\pi$ between the two types of particles,
the random visons serve as sources of $\pi$ fluxes, which lead to
quantum interference corrections to the diffusion of the spinons.

\section{Analysis of model-XX}
\label{sec:numerics_XX}

Here we present numerical results for the model-XX introduced in Sec.~\ref{sec:xxgauge-model},
organized in the same way as the numerical studies of model-X in Sec.~\ref{sec:tfimgauge}. Lanczos
ED calculations were carried out for $L=2$ systems. QMC results for larger systems up
to $L=6$ were again obtained using the SSE method supplemented  with quantum replica exchange.
We fixed $\Gamma_m = 1.0$ and simulated several replicas at different values of $J_x$ across the
two phases. The values are chosen such that the acceptance rate of swapping
neighboring replicas is in the range $0.4 \sim 0.6$. Our findings and arguments are
very similar to those for model-X discussed in Sec.~\ref{sec:tfimgauge}, with the exception of issues
pertaining to the ferromagnetic phase, and we therefore keep the
discussion brief in this section.

\subsection{Fidelity susceptibility}

\begin{figure}[!ht]
\includegraphics[width=75mm]{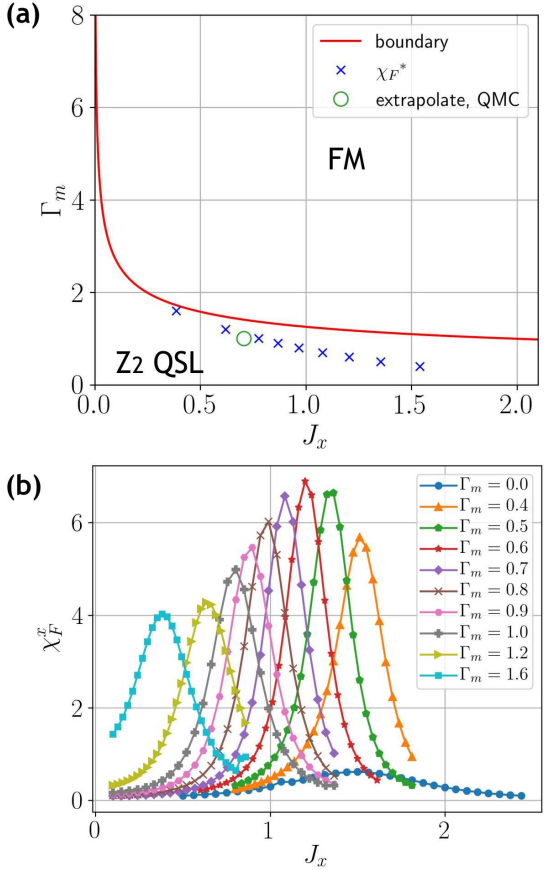}
\caption{(a) Phase diagram of the model with XX interactions. The red curve is the approximate phase
boundary obtained from the asymptotic $\Gamma_m \to \infty$ mapping to the TFIM. This boundary separates
the $\mathbb{Z}_2$ topological quantum spin liquid and the ferromagnetic phases. The blue crosses are points
on the boundary estimated from the maximum of the fidelity susceptibility in (b), calculated using ED on an $L=2$ system.
The green open circle indicates the $L\rightarrow \infty$ extrapolated transition point $J_x \approx 0.706$ from QMC
simulations at $\Gamma_m=1.0$.}
\label{Fig.Wl.Xf}
\end{figure}

As in the previous study of model-X, we first discuss Lanczos ED
results for the fidelity susceptibility $\chi^{x}_F$ defined as in
Eq.~\eqref{eq.Xf2} with the substitution $\Gamma_g \rightarrow
J_x$. In Fig.~\ref{Fig.Wl.Xf}(b), we show $\chi^x_F$ versus $J_x$ for
several values of $\Gamma_m$.  In all cases, we observe a peak
indicative of a phase transition. The locations of the maxima are
shown along with the perturbative phase boundary in
Fig.~\ref{Fig.Wl.Xf}(a). 

\begin{figure}[t]
\includegraphics[width=75mm]{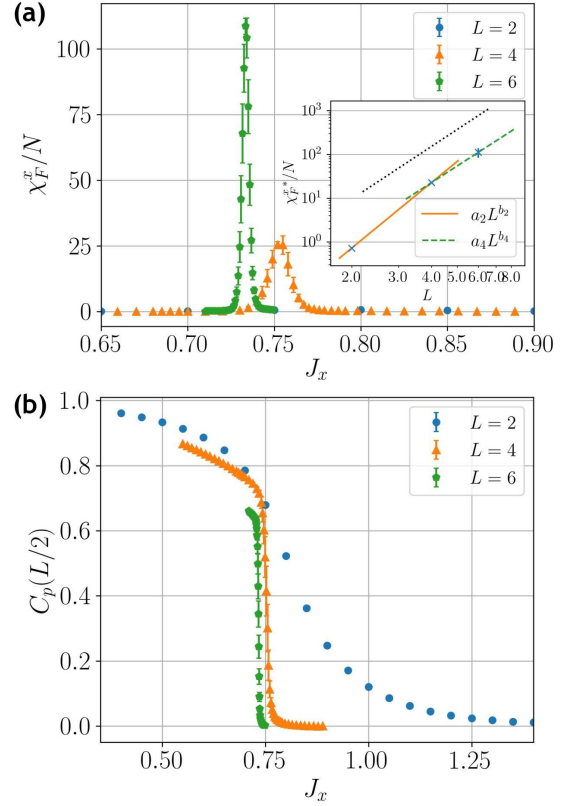}
\caption{(a) Fidelity susceptibility of model-XX at $\Gamma_m = 1.0$ for different system sizes $L$ calculated using SSE simulation at
inverse temperature $\beta = 4L$. The inset shows power-law fits $y \propto L^{b_L}$ to the  maximum values for system sizes $L$ and $L+2$.
The exponents are $b_2 \approx 4.9$ and $b_4 \approx 3.9$. The slope of dotted line corresponds to the predicted exponent $b_{\infty} = 2(d+1)-d = 4$
expected for a first-order transition. (b) QMC results at $\Gamma_m = 1.0$ for the Wilson loop correlation function, Eq.~\eqref{eq.Cp}.}
\label{Fig.Wl.XX}
\end{figure}

In Fig.~\ref{Fig.Wl.XX}(a) we plot SSE
results for larger systems. As expected we find a maximum that
diverges with increasing system size, showing a true phase boundary
and only two phases. Analyzing the peak height using the effective
exponent $b_L$ defined with system sizes $L$ and $L+2$, we find
$b_2 \approx 4.8$ and $b_4 \approx 3.9$. These exponents are again
significantly larger than the $2/\nu-d \approx 2/0.63-2 \approx 1.175$
for the (2+1)D Ising universality class, but close to
$b = 2(d+z)-d = 4$ for a first-order transition when $z=1$. We note
one difference with respect to the previous model, as seen in
Fig.~\ref{Fig.Wl}, in that case the exponent $b_L$ increases with $L$,
while in the present case it decreases. One may then question whether
the value $b=4$ is really obtained in the limit $L \to
\infty$. Nevertheless, all our complementary results to be presented
below also lend support to a first-order transition.
   
\subsection{Topological order}

To detect the topological order, we again consider the correlation function $C_p(L/2)$ between two parallel non-contractible Wilson loops, defined 
previously in Eq.~\eqref{eq.Cp}. Fig.~\ref{Fig.Wl.XX}(b) shows results at $\Gamma_m = 1.0$. Here a discontinuity reflecting the first-order transition
develops more clearly as compared to the results for model-X in Fig.~\ref{Fig.Wlb}, thus suggesting a more strongly first-order transition in this case. Note, however, that the parameter values chosen for the two models in these figures, $\Gamma_g=0.2$ and $\Gamma_m=1$, are not directly comparable. In both
cases, the strength of the discontinuity will vary with the model parameters.

\begin{figure}[t]
\includegraphics[width=\linewidth]{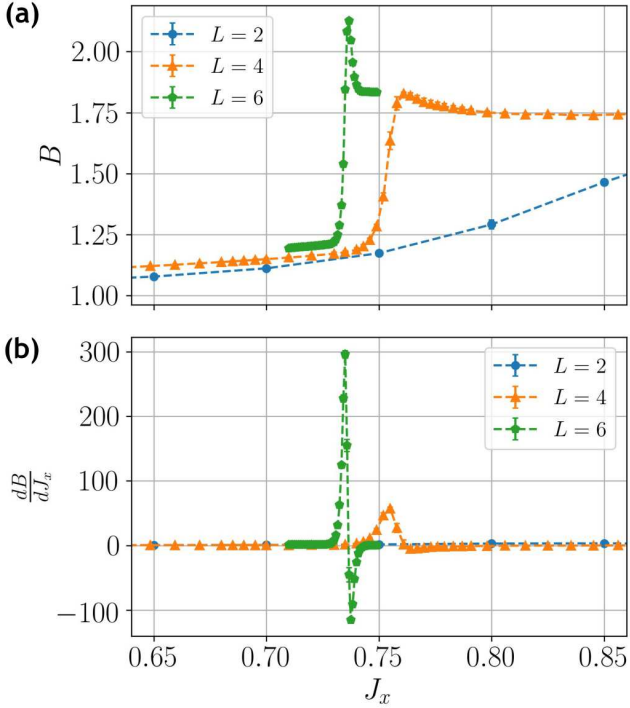}
\caption{Results for model-XX at $\Gamma_m = 1.0$. (a) The Binder ratio of the Wilson loop order parameter defined
with both components in Eq.~(\ref{pxydef}); $P^2 = P_x^2 + P_y^2$. The divergent peak next to the step indicates phase coexistence
at a first-order transition. Accordingly, in (b) the derivative of the Binder ratio shows divergent positive and negative peaks.}
\label{Fig.B2.XX}
\end{figure}

As in Sec.~\ref{sec:tfimgauge}, we use the Wilson loop order parameters $P_x = \sum_a p_{x,a}$ and $P_y = \sum_a p_{y,a}$ to confirm the extent
of the topological phase. We use the Binder ratio $B$ as defined in Eq.~\eqref{bratiodef_xy} with both components taken into account. The results, shown
in Fig.~\ref{Fig.B2.XX}(a) exhibit developing step functions with associated peaks indicative of a first-order transition. The derivatives exhibit the
expected divergent peaks. Because of the limited system sizes, we refrain from analyzing the peaks further. We have used the peaks to extrapolate the
transition point to infinite size and show the result with the green circle in the phase diagram in Fig.~\ref{Fig.Wl.Xf} at $\Gamma_m=1$. As in the model-X 
with transverse field on the gauge spins, we find only a small difference between the QMC result and the $L=2$ ED result in this case.
	   
\subsection{Energy derivative}

\begin{figure}[t]
\includegraphics[width=75mm]{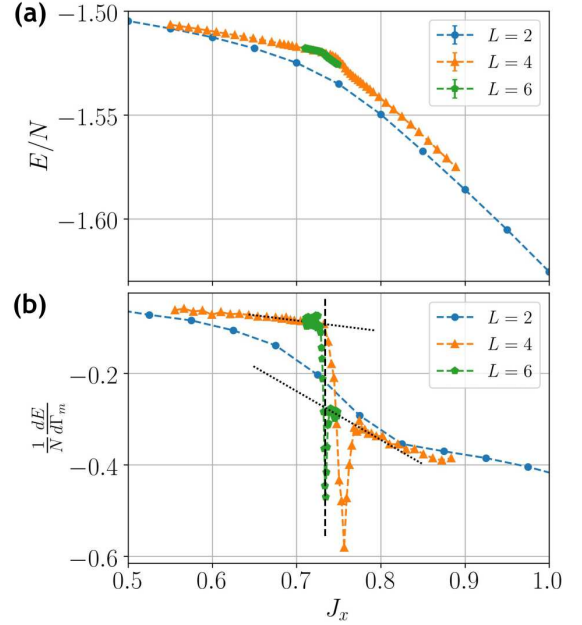}
\caption{(a) Energy per spin of model-XX obtained in the same simulations as the other quantities at $\Gamma_m = 1.0$ and
temperature $T=(4L)^{-1}$. (b) The derivatives are evaluated using the linear approximation using the data in (a). The dotted
lines are fits to the $L=4$ results away from the peak and demonstrate a jump in the energy derivative at the transition 
(here represented by the peak location as indicated by the vertical dashed line).}
\label{FigdExx}
\end{figure}

We present further evidence of a first-order phase transition from the energy density. As shown in Fig.~\ref{FigdExx}(a), in this case
we observe a clear kink behavior for the larger system sizes, $L\geq4$, and the derivative in Fig.~\ref{FigdExx}(b) accordingly shows a strong
discontinuity developing. 

\subsection{Level spectroscopy}

\begin{figure}[t]
\includegraphics[width=75mm]{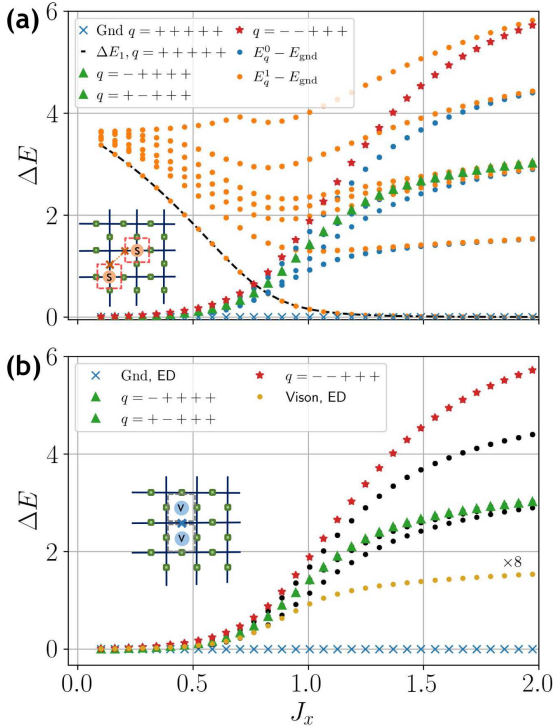}
\caption{Level spectrum relative to the ground state for the $L=2$ system with XX interactions, calculated with Lanczos ED. The organization of panels 
(a) and (b) is as in the corresponding Fig.~\ref{Fig.Vgap_tfim} for the model-X. We refer to the same for further explanation of the visualization of
the spectrum. The lowest spinon excitation, shown as orange circles with a dashed line, is the first excited state from the same block as the ground
state (which has $\mathbf{q} = +++++$). This state is antisymmetric with respect to spin inversion and forms the two-fold degenerate multiplet together
with the corresponding symmetric state in the ferromagnetic phase.}
\label{Fig.Vgap_XX}
\end{figure}

We have again used Lanczos ED to find low-lying states for each block
of quantum numbers characterizing the combinatorial $\mathbb{Z}_2$
symmetries in the $L=2$ system.  In Fig.~\ref{Fig.Vgap_XX}, we present
the two smallest gaps versus the XX coupling $J_x$ at $\Gamma_m =
1.0$. The lowest states in the sectors $G_x = \pm$, $G_y = \pm$ and
$G_i = +$ again are those that will eventually become degenerate as
$L\rightarrow \infty$ in the topological phase, and these states are
highlighted with different symbols. The lowest energy excitations in
the topological phase, states with visons, form levels very similar to
what we saw in the model-X.  However, the spinon
spectrum looks very different. Due to the $XX$ ferromagnetic interaction,
spinons created in pairs within one of the sublattices has lower energy
comparing to the one created in neighboring pairs (created by a single
$\sigma_x$ operation) as illustrated in Fig.~\ref{Fig.Vgap_XX}(a).

At $J_x \rightarrow 0$, only spinon excitations exist, with a gap size of order O($J$), as can be seen
in Fig.~\ref{Fig.Vgap_XX}(a) where these levels are marked in orange. The vison gap opens with increasing
$J_x$, as the effective model from the lowest order in perturbation takes the form
\begin{align}
H_{\rm pert} = \sum_{s} H_s - g\sum_p G_p.
\end{align}
Here $G_p$ is the local gauge generator (plaquette term) that appears at
10th order, with $g \sim \Gamma_m^8J_x^2/J^9$, which should be compared to
the 12th order perturbative Hamiltonian in the case of model-X.

In the case of the XX interaction used here, there is an additional
gauge spin inversion symmetry in $x$-basis that is not present in the
model-X. Define the
inversion operator $V = \prod_i \sigma_i^{z}$ as the product of all
$\sigma^z$ gauge-spins. This operator clearly commute with Hamiltonian
and its quantum numbers $v = \pm$ correspond to symmetric or
antisymmetric states. We find that all the lowest energy levels of
the 32 gauge blocks (blue) are symmetric and the second state (orange) is always
anti-symmetric except for the highest energy level shown in Fig.~\ref{Fig.Vgap_XX}(a), which exhibits an actual level crossing
(see appendix ~\ref{app:XX_V} for further details).  

Among all the spinon excitations, the lowest one
belongs to the same sector as the ground state, with $\mathbf{q} =
+++++$. This excitation, which is
marked with a dashed line in Fig.~\ref{Fig.Vgap_XX}(a), becomes
degenerate with the ground state for large $J_x$, reflecting the
ferromagnetic Ising order with spontaneously broken $\mathbb{Z}_2$
symmetry in the thermodynamic limit.

For the vison excitations, since $\sigma^z$ operators do not commute
with local gauge operators, we can identify the vison excited states
simply by considering the quantum number blocks that couple to the
ground state through the on-site $\sigma^z$ operators. In
Fig.~\ref{Fig.Vgap_XX}(b), we plot ED results for the same parameters
as in Fig.~\ref{Fig.Vgap_XX}(a), but with a change in scale to focus
on the vison states. These states are gapped for all $J_x > 0$, but
the gaps are much smaller in the topological phase than in the FM
phase. The lowest vison state, which is marked by yellow symbols in
Fig.~\ref{Fig.Vgap_XX}(b), contains two $m$ particles, and is
eight-fold degenerate on the small system considered here. The other
levels marked by blue in Fig.~\ref{Fig.Vgap_XX}(a) correspond to
states with more than two (an even number of) visons.

\section{Conclusions and Discussion}
\label{sec:conclusion}

We have presented a numerical study of spin models with only one- and
two-spin interactions that realize a combinatorial $\mathbb{Z}_2$
gauge symmetry. We considered two models that only differ by the
kinetic terms given to the gauge spins: model-X (containing a
transverse field) and model-XX (containing $XX$ interactions). We
found conclusive evidence for an extended $\mathbb{Z}_2$ topological
quantum spin liquid phase in both models.

In the case of model-X, we identified two phases; a topological phase
and a paramagnetic phase. We demonstrated a first-order quantum phase
transition between these phases, in contrast to the well known continuous
transition of the conventional $\mathbb{Z}_2$ lattice gauge model. In
model-XX we identified a topological phase and a competing
ferromagnetic state. Our data also support a first-order transition
between these two phases in model-XX. Perturbatively, the $XX$
interaction of model-XX generates a plaquette operator $G_p$ at a
lower order in perturbation theory as compared to the transverse field
of model-X, and therefore the size of the vison gap increases, as we
also observe.

The presence of the first-order transition between
the topological and the competing state, in both models, raises the
following interesting question: As we have discussed in the paper,
in the limit of a large transverse field $\Gamma_m$ on the matter
spins, the models map to the usual Ising gauge model, which has a
continuous transition. An important question is then whether the
continuous transition persists for some finite range of values of
$\Gamma_m$, or whether it turns first-order immediately. This
question can in principle be answered by considering the corrections
to the usual $\mathbb{Z}_2$ gauge model in Eq.~(\ref{hz2gaugemodel}), which
will appear when carrying out a perturbative expansion to higher
order in $\Gamma_m^{-1}$. The question is then whether these
corrections are renormalization-group relevant or irrelevant at the
critical point. While we have not carried out this expansion and
duality mapping, it appears likely that the additional interactions
generated in the Ising model will involve products of more than two
spins, and most likely these interactions will be irrelevant at the
Ising critical point. Thus, we suspect that there will be indeed a
tricritical point separating continuous Ising transitions and
first-order transitions for large values of $\Gamma_m$ in
Figs.~\ref{Fig.Xf} and \ref{Fig.Wl.Xf}. We leave tests of this
hypothesis open for future work.

\begin{acknowledgments}

 K.-H. W. and C.C. are supported by DOE Grant No. DE-SC0019275. 
 Z.-C.Y. acknowledges funding by the NSF PFCQC program.
 A.W.S. is supported by Simons Investigator Grant.~No.~511064.  
 The numerical
 simulations were carried out on the Shared Computing Cluster managed
 by Boston University's Research Computing Services.
\end{acknowledgments}

\appendix

\section{$\mathbb{Z}_2$ gauge symmetry and conserved quantum numbers}
\label{app:Z2sym}
		
Recall that in the $\mathbb{Z}_2$ lattice gauge theory,
\begin{align}
  H_{\mathbb{Z}_2} = J\sum_i A^z_i - h_x \sum_i \sigma_x^{i} 
\end{align}
where $A^z_i$ is the star operator defined as  $A^z_i = \sigma_z^{1}\sigma_z^{2}\sigma_z^{3}\sigma_z^{4}$
acting on 4 spins emanating from a single site as shown in Fig.~\ref{fig.mapping}(b).
The local gauge generator $G = \prod \sigma_x$ is defined as a product
of $\sigma_x$ operators around an elementary
plaquette [shown as the blue cross in Fig.~\ref{fig.mapping}(b)],
which is a conserved quantity of the system, i.e.
$G\ket{E_{q}} = q\ket{E_{q}}$ where $\ket{E_{q}}$ is an energy
eigenstate.  Thus we can use these operators to characterize the
quantum number sectors. 
	
In analogy to the standard $\mathbb{Z}_2$ lattice gauge theory, in our model the 4-body interaction term $A_i$ is effectively generated with the term (See Ref.~\cite{chamon2019emulate} for further details)
\begin{align}
  J \sum_{a \in s} \left(\sum_{j \in s} W_{aj} \sigma^z_j \right) \mu^z_a \rightarrow -\gamma - \lambda\sum_{i\in s}\sigma_i^{z},
\end{align}  
and the monomial transformation leads to the modification of the local
gauge generator
\begin{equation}
  G_p = \prod_{s \in p}\mathcal{L}_s^{\mu}\prod_{i\in p} \sigma_i^{x},
\end{equation}
with the additional $\mathcal{L}_s^{\mu}$ representing the action of
monomial matrices $\mathcal{L}$ acting on matter spins.  All the
formulations that characterize the symmetry and quantum numbers in the
standard $\mathbb{Z}_2$ lattice gauge theory can also be applied to
our model with combinatorial $\mathbb{Z}_2$ gauge symmetry.
	
\begin{figure}[t]
\includegraphics[width=\linewidth]{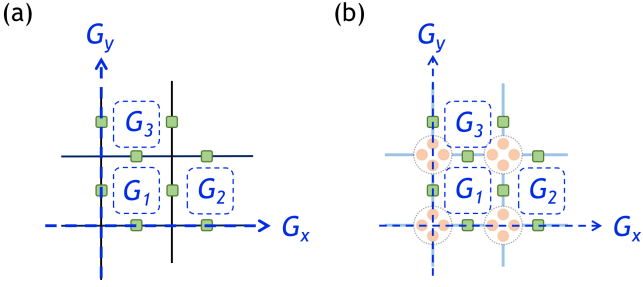}
\caption{(a) Definitions of $M$ independent gauge operators in the conventional $\mathbb{Z}_2$ model for a system of size $2\times2$.
(b) The corresponding operators define on the lattice of combinatorial $\mathbb{Z}_2$ model. Here, $G_x$ and $G_y$ are the gauge operators
defined on non-contractible loops along x and y direction respectively.}
\label{fig.gtype}
\end{figure}

There are in total $M = 2 + (L_xL_y - 1)$ such independent operators in a $L_x \times L_y$ system with periodic boundary conditions.
Within the $M$ operators, two of them are defined as a product of $\sigma_x$, $G_i \equiv \prod_j^{(i)} \sigma_x^{j}$ along non-contractible
loops where their quantum numbers uniquely characterize the 4-fold topological degeneracies of the ground state in the thermodynamic limit.
As shown in Figure.~\ref{fig.gtype}(a), $G_y$ is defined along a non-contractible loop in the $y$-direction, and $G_x$ is defined in the
$x$-direction. Other operators are local, defined as a product of $\sigma_x$ around an elementary plaquette of 4 spins. 

Using the fact that $G$ commutes with the Hamiltonian, we can construct eigenstates of the $G$ operators: $G\ket{q} = q\ket{q}$ where $q = \pm1$.
It is straightforward to see that for each $G$ operator, the eigenstate can be constructed by starting from a classical configuration (which we refer
to as the ``representative'' state \cite{Sandvik2010}) via $\ket{q} = (1 + qG)\ket{\mathrm{rep}}$. Since the Hamiltonian commutes simultaneously
with all $M$ operators, the state should be constructed with a product of $1+q_iG_i$ for all $G_i$.	
As an example, consider a $2 \times 2$ system as shown in Fig.~\ref{fig.gtype}, we have
\begin{align}
\ket{\mathbf{q}_i} = \frac{1}{Z}(1+q_xG_x)(1+q_yG_y)\prod_{j=1}^3 (1+q_jG_j)\ket{\mathrm{rep}_i}
\end{align}
where $\mathbf{q}$ indicates the quantum number set $\mathbf{q} = (q_x,q_y,q_1,q_2,q_3) = (\pm,\pm,\pm,\pm,\pm)$, $Z$ is the normalization factor and
subscript $i$ indicates the $i$-th state within the block. There are in total $2^M$ symmetry blocks in the system.
    
\section{SSE derivative of the Wilson loop Binder ratio} 
\label{app:dB2}

The derivative of the Binder ratio can be evaluated directly in SSE simulations, using the estimator derived here. Consider the Hamiltonian 
$H = JH_0 + \delta H_\delta$, where $\delta$ is the tuning parameter and $[H_0,H_\delta] \neq 0$. For any arbitrary diagonal observable $O$, the expectation 
can be express in the SSE representation as  \cite{Sandvik2010}
\begin{align}
\left<O\right> &= \frac{1}{Z}\sum_{\{\alpha_i\}} F(\beta,n) O_{\alpha_0\alpha_0}H^{\alpha_0\alpha_1} \cdots H^{\alpha_{M-1}\alpha_0} \nonumber\\
&=\frac{1}{Z}\sum_{\{\alpha_i\}} F(\beta,n)O_{\alpha_0\alpha_0}J^{n_J(\{\alpha_i\})}\delta^{n_\delta(\{\alpha_i\})} R\left[H_{a_i,b_i}\right]
\end{align}
where 
\begin{align}
F(\beta,n) \equiv \frac{\beta^n(M-n)!}{M!}, 
\end{align}
and we have used the short-hand notation $H^{\alpha_i\alpha_j} = \left<\alpha_i|H|\alpha_j\right>$ and $O_{\alpha_0\alpha_0} = \left<\alpha_0|O|\alpha_0\right>$. 
Further, $Z$ is the partition function, $M$ is the operator string length and $n$ is the number of non-identity operators in the current string. 
The quantity denoted $R\left[H_{a_i,b_i}\right]$ stands for the product of local Hamiltonian operators, $\prod_i H_{\alpha_i}$, where $H_{\alpha_i} = H_0^{i}$ 
or $H_\delta^{i}$.

The derivative of the observable with respect to the tuning parameter $\delta$ can be calculated from
\begin{align}
\frac{\partial\left<O\right>}{\partial\delta}
&= \frac{\left<On_\delta\right>}{\delta} - \left<O\right>\frac{\left<n_\delta\right>}{\delta} \nonumber\\
&= \frac{\left<On_\delta\right>-\left<O\right>\left<n_{\delta}\right>}{\delta},
\end{align}
where $n_{\delta}$ is the number of the $\delta$ operators in the string. We are interested in the Binder ratio of the Wilson loop order parameter, 
defined as in Eq.~(\ref{bratiodef_xy}). Using the above expressions we obtain
\begin{align}
\frac{\partial B}{\partial \delta} &=  \frac{\partial_\delta\left<P^4\right>}{\left<P^2\right>^2} - \frac{2\left<P^4\right>\left<P^2\right>\partial_\delta \left<P^2\right>}{\left<P^2\right>^4}  \nonumber\\
&= \frac{1}{\delta}\left[\frac{\left<P^4 n_\delta\right> +  \left<P^4\right>\left<n_\delta\right>}{\left<P^2\right>^2} - \frac{2\left<P^4\right>\left<P^2n_\delta\right> }{\left<P_x^2\right>^3} \right].
\end{align}
Here $P^2=p_x^2+P_y^2$ defined in Sec.~\ref{sec:topo1} is an equal-time quantity evaluated at a given ``time slice'' in the SSE configuration.

\section{Extraction of the vison gap} 
\label{app:gapXX}

To extract the vison gap from SSE simulations, we evaluate the imaginary-time correlator defined as 
\begin{align}
	G(\tau,\beta) = 1/N_g \sum_{i \in g} \left< \sigma_z^{i}(0)\sigma_z^{i}(\tau)e^{-\beta\hat{H}} \right>
\end{align} 
where $\sigma_z$ is the Pauli-$z$ operator acting on the gauge spin,
$\tau$ is the imaginary time and $\beta = 1/T$ is the inverse
temperature. The estimator is averaged over all the
$N_g = L \times L \times 2$ gauge spins. The operator $\sigma_z$
acting on the ground state creates a pair of visons, thus, at a
sufficiently low temperature, the extracted gap from the exponential
fitting of Green's function gives the estimation of the vison gap.

In order to obtain the gap correctly, it is essential that the
temperature is sufficiently low in the simulation. To elaborate on
this point, consider a finite temperature Green's function in
the basis of energy eigenstates
\begin{align}
  G(\tau,\beta)_i
  &\equiv \frac{1}{Z}\left< \sigma_z^{i}(0)\sigma_z^{i}(\tau)e^{-\beta\hat{H}} \right>
    \nonumber\\
  &= \frac{1}{Z}\sum_{a,b} \bra{a} \sigma_z^{i} e^{-\tau E_a} \ket{b}\bra{b} \sigma_z^{i} e^{-(\beta-\tau)E_b } \ket{a} \nonumber\\
  &= \frac{1}{Z}\sum_{a,b} |\bra{a}\sigma_z^{i}\ket{b}|^{2}e^{-\beta E_b} e^{-\tau(E_a - E_b)}.
\end{align}
For the few leading terms in a system at a sufficient low temperature we have 

\begin{align}
  G(\tau,\beta)
  &\approx \frac{e^{-\beta E_0}}{Z} \sum_{a}|\sigma^{z,i}_{a0}|^2 e^{-\tau\Delta_{a0}} \nonumber\\
  &+\frac{e^{-\beta E_1}}{Z}  \sum_{a}|\sigma^{z,i}_{a1}|^2e^{-\tau\Delta_{a1}} \nonumber \\
  &= \frac{e^{-\beta E_0}}{Z} \left[|\sigma^{z,i}_{10}|^2 e^{-\tau\Delta_{10}}  + \sum_{a=2} |\sigma^{z,i}_{a0}|^2 e^{-\tau\Delta_{a0}}\right] \nonumber\\
  &+\frac{e^{-\beta E_1}}{Z}  \sum_{a}|\sigma^{z,i}_{a1}|^2e^{-\tau\Delta_{a1}},
\end{align}
where $\Delta_{ab} = E_a - E_b$ is the energy difference between the two states $a$ and $b$, and we ignore all the terms with $a=b$ since the diagonal matrix 
element $\sigma^{zz,i}_{aa}$ vanishes. We further separate the dominant terms by rewriting the above expression as
\begin{align}
  G(\tau,\beta)
  &\approx \frac{e^{-\beta E_0}}{Z}|\sigma^{zz,i}_{10}|^2 e^{-\tau\Delta_{10}} \left[1+\sum_{a=2} \frac{|\sigma^{zz,i}_{a0}|^2}{|\sigma^{zz,i}_{10}|^2}r^{-\tau \Delta_{a1}}\right] \nonumber\\
  &+ \frac{e^{-\beta E_1}}{Z}\left[|\sigma^{zz,i}_{01}|^2e^{\tau \Delta_{10}} + \sum_{a=2}|\sigma^{zz,i}_{a1}|^2e^{-\tau\Delta_{a1}}\right].
\end{align}
Considering only the leading three terms related to the gap $\Delta_{10}$ we are interested in, and with the fact that 
$|\sigma_{10}^{zz,i}|^2 = |\sigma_{01}^{zz,i}|^2$, we have
\begin{align}
  G(\tau,\beta)
  &\approx \frac{e^{-\beta E_0}}{Z}|\sigma^{zz,i}_{10}|^2 e^{-\tau\Delta_{10}} \left[1+\sum_{a=2} \frac{|\sigma^{zz,i}_{a0}|^2}{|\sigma^{zz,i}_{10}|^2}r^{-\tau \Delta_{a1}}\right] \nonumber\\
  &+ \frac{e^{-\beta E_1}}{Z}|\sigma^{zz,i}_{10}|^2e^{\tau \Delta_{10}} \nonumber\\
  &= |\sigma^{zz,i}_{10}|^2 \frac{e^{-\beta E_0}}{Z} \times \nonumber \\
  & \{e^{-\tau\Delta_{10}}\left[1 + \sum_{a=2} \frac{|\sigma^{zz,i}_{a0}|^2}{|\sigma^{zz,i}_{10}|^2}r^{-\tau \Delta_{a1}} \right]  + e^{-(\beta-\tau)\Delta_{10}}\}
\end{align}

\begin{figure}[t]
\includegraphics[width=0.5\linewidth]{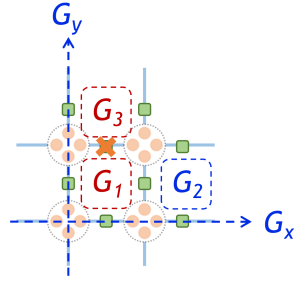}
\caption{The on-site $\sigma_z$ operator creates a pair of visons when acting on the ground state with quantum number $\mathbf{q} = (+,+,+,+,+)$.  
The quantum number corresponding to the operators $G_1$ and $G_3$ is then changed, leading to an excited state with quantum number $\mathbf{q}' = (+,+,-,+,-)$. }
\label{fig.change_q}
\end{figure} 

Furthermore, notice that the $\sigma_z$ operator does not commute with
the gauge operator $G_p$. In fact, if we operate on the site with spin
$\sigma_z^{1}$ with the gauge operator
$G_p = \sigma_x^{1}\sigma_x^{2}\sigma_x^{3}\sigma_x^{4}$, the
$\sigma_z$ operator changes the quantum number corresponding to $G_p$
since
\begin{align}
  \sigma_z^{1}G_p = -G_p\sigma_z^{1}.
\end{align}
As illustrated in Fig.~\ref{fig.change_q}, acting with a $\sigma_z$ operator on the ground state with quantum number set $\mathbf{q} = (+,+,+,+,+)$ creates a 
pair of visons and thereby changes the quantum numbers associated with $G_1$ and $G_3$, leading to a new quantum number set $\mathbf{q}' = (+,+,-,+,-)$. 
This means that the ground state will have non-zero matrix elements only to the states with the right quantum number set. Thus, we can safely assume
the matrix elements $|\sigma_{a0}^{zz}|^2 = 0$ for low levels. If we ignore these term in the summation we have
\begin{align}
G(\tau,\beta) \sim \{e^{-\tau\Delta_{10}}  + e^{-(\beta-\tau)\Delta_{10}}\} \label{eq.G}.
\end{align}
In our simulations, we evaluate this non-equal time correlator at various values of $\tau$ and extract the gap $\Delta_{10}$ by fitting the results 
to Eq.~\eqref{eq.G}.

\begin{figure}[t]
  \includegraphics[width=75mm]{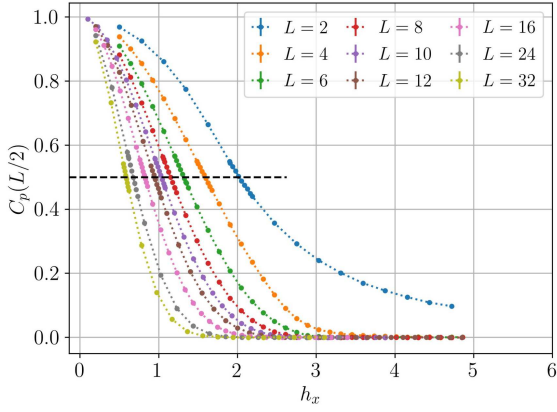}
  \caption{Wilson loop correlator calculated with SSE simulations of the
    square-lattice TFIM for several system sizes at inverse temperature
    $\beta = 2L$. The inset shows the finite-size scaling
    of the location $h_x^*(L^{-1})$ for which the value of the
    correlator is $1/2$, as indicated by the horizontal dashed line.
    A power-law fit $y = a + bL^{-c}$ for the $L\geq 12$ data
    gives $a = -0.02(1)$, $b = 3.15(2)$ and $c = 0.472(7)$, indicating
    a vanishing value of the correlator at any transverse field
    in the thermodynamic limit.}
  \label{fig.app.pipjtfim2d}
\end{figure}

\begin{figure}[t]
  \includegraphics[width=65mm]{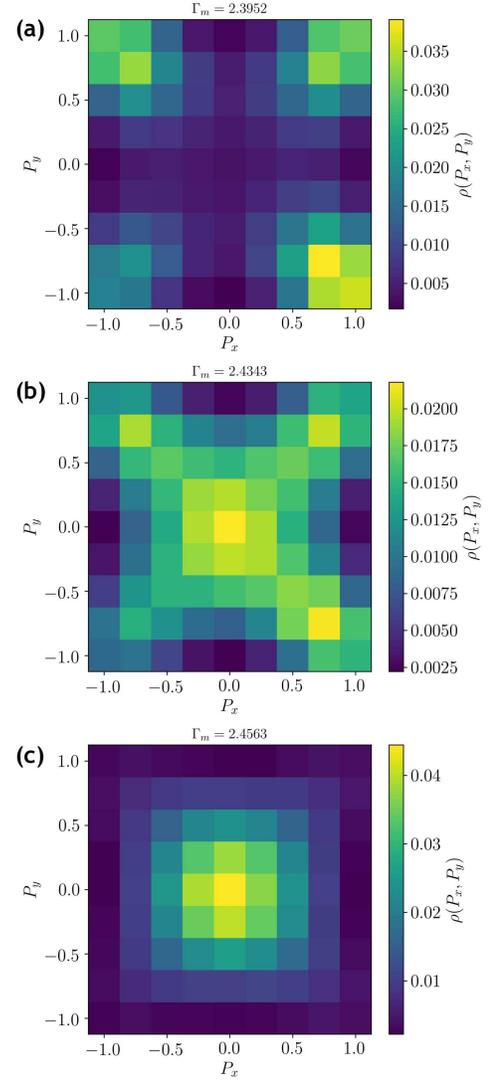}
  \caption{Distribution $\rho(P_x,P_y)$ of the Wilson loop order parameter accumulated in SSE simulation at $\Gamma_g=0.2$ and three
    different values of the matter field; (a) $\Gamma_m = 2.3952$ (in the $\mathbb{Z}_2$ QSL phase), (b) $\Gamma_m = 2.4343$ (close to the
    transition point), and (c) $\Gamma_m = 2.4563$ (in the confined phase). Near the transition point, in (b), five peaks are clearly observed,
    reflecting phase coexistence at a first-order transition. }
  \label{fig.app.pxpyL8}
\end{figure}

\section{Parallel Wilson loop operator in square lattice transverse field Ising model} \label{app:2DTFIM}

To show that the parallel Wilson loop operator does not detect the
long range ferromagnetic phase, we perform simulations of the TFIM
on $L \times L$ square lattices with periodic boundary conditions at
inverse temperature $\beta = 2L$. We compute the Wilson loop correlator
defined in Eq.~(\ref{eq.Cp}). As shown in Fig.~\ref{fig.app.pipjtfim2d},
we draw the horizontal line at $C_P=1/2$  and extract the corresponding
value $h_x^*(L)$ of the transverse field. In the inset of Fig.~\ref{fig.app.pipjtfim2d},
we demonstrate finite-size scaling of the value $h_x^*$ as a function of $1/L$ with a power-law
fit of the form $y = a+bL^{-c}$. Including data for the four largest system
sizes, $L \geq 12$, we find the best fit with $a = -0.02(1)$, $b = 3.15(2)$
and $c = 0.472(7)$, confirming the expectation Wilson loop correlator
vanishes for a conventional FM phase in the thermodynamic limit
at any finite transverse field.

\section{Wilson loop order parameter distribution and phase coexistence} \label{app:TFpxpy}

\begin{figure}[t!]
  \includegraphics[width=75mm]{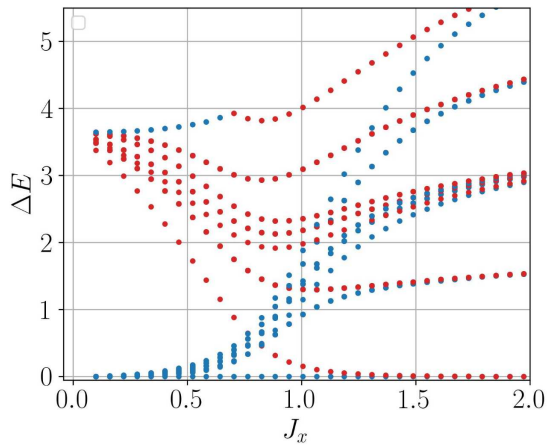}
  \caption{The low-energy levels previously shown in Fig.~\ref{Fig.Vgap_XX}, now marked by the eigenvalues $\pm 1$ of the
    spin-inversion operator, Eq.~(\ref{vopdef}), in model-XX at $\Gamma_m=1$. The gaps to the symmetric and antisymmetric states are marked
    with blue and red symbols, respectively. Out of all 64 states shown here, all the vison excitation states as well as the ground
    state are symmetric. The spinon excitations are all antisymmetric, except for the highest one, where a level
    crossing causes a change in symmetry at $J_x \approx 0.7$.}
  \label{fig.app.vxx}
\end{figure}

Here, we present simulation results for Wilson loop order parameter
distribution $\rho(P_x,P_y)$ for model-X. The two components of
the order parameter are defined as in Eq.~(\ref{pxydef}).
Fig.~\ref{fig.app.pxpyL8} shows color-coded plots of the
distribution as the phase transition is traversed. Near the
transition point, in Fig.~\ref{fig.app.pxpyL8}(b), we observe a
five-peak structure, indicating phase coexistence at a first-order
transition. In (a) and (c) we observe distributions expected in systems
with and without topological order, respectively.

Note that the four peaks at the corners in Fig.~\ref{fig.app.pxpyL8}(a)
and Fig.~\ref{fig.app.pxpyL8}(b) should mathematically be of equal size,
but they differ here because of the slow migration of the simulation 
between these peaks, which are separated by tunneling barriers in the
SSE configuration space.

\section{Additional $\sigma_x$ spin inversion symmetry in model-XX} \label{app:XX_V}

Model-XX possesses a spin inversion symmetry corresponding to the operator
\begin{align}
  V \equiv \prod_i \sigma_i^{z}
          \;,
\label{vopdef}
\end{align}
which commutes with the Hamiltonian. In Fig.~\ref{fig.app.vxx} we show the
same energy levels of the model-XX at $\Gamma_m = 1.0$ as previously in
Fig.~\ref{Fig.Vgap_XX}. Here different colors indicate symmetry or
antisymmetry with respect to $V$. In the FM phase at large $J_x$ the gauge
spins order along spin-$x$ direction, and the first excited state and the
ground state are both from the block with topological quantum number set
$\mathbf{q}=(+++++)$. In Fig.~\ref{Fig.Vgap_XX} we can observe how the symmetric
and antisymmetric $V$ states become degenerate (strictly in the thermodynamic
limit) to allow spontaneous symmetry breaking in the FM phase.

\newcommand{\npb}{Nucl. Phys. B}\newcommand{\adv}{Adv.
	Phys.}\newcommand{\RMP}{Rev. Mod. Phys.}\newcommand{\PRB}{Phys. Rev.
	B}\newcommand{\PR}{Phys. Rev.}\newcommand{\PRL}{Phys. Rev.
	Lett.}\newcommand{\plb}{Phys. Lett.
	B}\newcommand{\jstat}{JSTAT}\newcommand{\JPC}{J. Phys,
	C}\newcommand{\njp}{New Journal of Physics}

\end{document}